\documentclass[fleqn,usenatbib]{mnras}

\usepackage[T1]{fontenc}

\usepackage{ae,aecompl}
\usepackage[utf8]{inputenc}
\usepackage{hyperref}

\usepackage{graphicx}	
\usepackage{amsmath}	
\usepackage{amssymb}	
\usepackage{ulem}	
\usepackage{booktabs, blkarray}
\usepackage{subcaption} 
\usepackage{newtxtext,newtxmath}

\pdfoutput=1 

\newcommand{\hmpc }{$h^{-1}$Mpc}
\newcommand{\hgpc }{$h^{-1}$Gpc}
\newcommand{\oii}{$[\textnormal{O}_\textsc{II}]$} 
\newcommand{\halpha}{$\textnormal{H}\alpha$} 
\newcommand{\EZ}{\textsc{EZmocks}}
\newcommand{\NSeries}{\textsc{Nseries}}
\newcommand{\OR}{\textsc{OuterRim}}

\title[RSD analysis with eBOSS DR16 voids]{The Completed SDSS-IV Extended Baryon Oscillation Spectroscopic Survey: Growth rate of structure measurement from cosmic voids}

\author[eBOSS Collaboration]{\parbox{\textwidth}{
Marie Aubert$^1$\thanks{maubert@ipnl.in2p3.fr}, 
Marie-Claude Cousinou$^{1}$, 
St\'ephanie Escoffier$^{1}$,
Adam J. Hawken$^{1}$,
Seshadri Nadathur$^{2}$, 
Shadab Alam$^{3}$,
Julian Bautista$^{2}$, 
Etienne Burtin$^{4}$, 
Chia-Hsun Chuang$^{5}$,
Axel de la Macorra$^{6}$,
Arnaud de Mattia$^{4}$,
H\'ector Gil-Mar\'in$^{7,8}$,
Jiamin Hou$^{9}$, 
Eric Jullo$^{10}$,
Jean-Paul Kneib$^{11}$,
Richard Neveux$^{4}$,
Graziano Rossi$^{12}$,
Donald Schneider$^{13}$,
Alex Smith$^{4}$,
Am\'elie Tamone$^{11}$,
Mariana Vargas Maga\~na$^{6}$,
Cheng Zhao$^{11}$
} \vspace*{4pt} \\ 
$^{1}${Aix Marseille Univ, CNRS/IN2P3, CPPM, Marseille, France}\\
$^{2}${Institute of Cosmology and Gravitation, University of Portsmouth, Burnaby Road, Portsmouth, PO1 3FX, UK}\\
$^{3}${Institute for Astronomy, University of Edinburgh, Royal Observatory, Edinburgh, EH9 3HJ, UK}\\
$^{4}${IRFU, CEA, Universit\'e Paris-Saclay, F-91191 Gif-sur-Yvette, France} \\
$^{5}${Kavli Institute for Particle Astrophysics and Cosmology, Stanford University, 452 Lomita Mall, Stanford, CA 94305, USA}\\
$^{6}${Instituto de F\'isica, 
Universidad Nacional Aut\'onoma de M\'exico, 
Apdo. Postal 20-364, Ciudad de M\'exico, M\'exico}\\ 
$^{7}${Institut de Ci\'encies del Cosmos,  Universitat  de  Barcelona,  ICCUB,  Mart\'i  i  Franqu\'es  1,  E08028  Barcelona,  Spain}\\
$^{8}${Institut  d'Estudis  Espacials  de  Catalunya  (IEEC),  E08034  Barcelona,  Spain}\\
$^{9}${Max-Planck-Institut f\"ur Extraterrestrische Physik, Postfach 1312, Giessenbachstrasse 1, 85748 Garching bei M\"unchen, Germany}\\
$^{10}${Aix Marseille Univ, CNRS, CNES, LAM, Marseille, France}\\
$^{11}${Institute of Physics, Laboratory of Astrophysics, \'Ecole Polytechnique F\'ed\'erale de Lausanne (EPFL), Observatoire de Sauverny, CH-1290 Versoix, Switzerland}\\
$^{12}${Department of Physics and Astronomy, 
Sejong University, Seoul, 143-747, Korea}\\
$^{13}${Institute for Gravitation and the Cosmos, Pennsylvania State University, University Park, PA 16802, USA}\\
}

\date{Accepted XXX. Received YYY; in original form ZZZ}

\pubyear{2020}

\begin{document}
\label{firstpage}
\pagerange{\pageref{firstpage}--\pageref{lastpage}}
\maketitle

\begin{abstract}
We present a void clustering analysis in configuration-space using the completed Sloan Digital Sky Survey IV (SDSS-IV) extended Baryon Oscillation Spectroscopic Survey (eBOSS) DR16 samples. These samples consist of Luminous Red Galaxies (LRG) combined with the high redshift tail of the SDSS-III Baryon Oscillation Spectroscopic Survey (BOSS) DR12 CMASS galaxies (called as LRG+CMASS sample), Emission Line Galaxies (ELG) and quasars (QSO). We build void catalogues from the three eBOSS DR16 samples using a ZOBOV-based algorithm, providing 2,814 voids, 1,801 voids and 4,347 voids in the LRG+CMASS, ELG and QSO samples, respectively, spanning the redshift range $0.6<z<2.2$. 
We measure the redshift space distortions (RSD) around voids using the anisotropic void-galaxy cross-correlation function and we extract the distortion parameter $\beta$. 
We test the methodology on realistic simulations before applying it to the data, and we investigate all our systematic errors on these mocks. We find $\beta^{\rm LRG}(z=0.74)=0.415\pm0.087$,  $\beta^{\rm ELG}(z=0.85)=0.665\pm0.125$ and $\beta^{\rm QSO}(z=1.48)=0.313\pm0.134$, for the LRG+CMASS, ELG and QSO sample, respectively. The quoted errors include systematic and
statistical contributions. 
In order to convert our measurements in terms of the growth rate $f\sigma_8$, we use consensus values of linear bias from the eBOSS DR16 companion papers, resulting in the following constraints: $f\sigma_8(z=0.74)=0.50\pm0.11$, $f\sigma_8(z=0.85)=0.52\pm0.10$ and $f\sigma_8(z=1.48)=0.30\pm0.13$. 
Our measurements are consistent with other measurements from eBOSS DR16 using conventional clustering techniques. 
\end{abstract}




\section{Introduction}
Observational cosmology has been leading for more than 20 years now to the discovery of one of the greatest puzzles in contemporary physics: the acceleration of cosmic expansion. Discovered in 1998 through the study of type Ia supernovae~\citep{Perlmutter99,Riess1998}, cosmic acceleration can be understood as a repulsive effect counteracting gravitational attraction, often depicted as a dark energy which is encoded by the cosmological constant $\Lambda$. 
In an attempt to find the underlying theory behind the late-time cosmic acceleration, two widely accepted approaches are generally proposed. The first is to assume the presence of an additional degree of freedom in the form of scalar fields as a way to allow the dark energy to evolve~\citep{Copeland2006}. The second is to consider modified gravitational theories which deviate from Einstein's General Relativity (GR) on cosmological scales~\citep{Nojiri2017}. 

To break the degeneracy between dark energy and modified gravity, a key test is to measure the linear growth rate of structure, which provides a measure of how fast structure is assembled in the Universe as a function of cosmic time.  
Constraints on the growth rate can be provided by galaxy redshift surveys. Indeed, galaxies that trace cosmic structure are subject to peculiar velocities which add an additional Doppler component to the cosmological redshift due to the Hubble flow. This line-of-sight component introduces anisotropies in the inferred spatial clustering of galaxies, a signal known as redshift space distortions (RSD)~\citep{Kaiser1987}. Since these velocities are related to the gravity of the cluster, the RSD pattern can be used to extract information on the growth rate, and thus allows us to distinguish between different theories of gravity~\citep{Peacock2001,Guzzo2008}. In GR, the growth rate is well approximated by the empirical relationship~\citep{Linder2005}:
\begin{eqnarray}
f = \Omega_m^{\gamma},
\end{eqnarray}
where $\Omega_m$ is the matter density and $\gamma=0.55$.

Techniques for extracting the RSD signal from galaxy redshift surveys have developed considerably over the past decade~\citep{Guzzo2008}, in particular from large datasets such as the 6 degree Field Galaxy Survey 6dFGS~\citep{Beutler2012}, the WiggleZ Dark Energy Survey~\citep{Blake2011,Contreras2013}, the VIMOS Public Extragalactic Redshift Survey (VIPERS)~\citep{Pezzotta2017,delaTorre2016}, the Baryon Oscillation Spectroscopic Survey (BOSS)~\citep{Alam2017}, the Subaru FMOS galaxy redshift survey (FastSound)~\citep{Okumura2016}, and recently the extended-BOSS DR14~\citep{Gil-Marin2018,Zarrouk2018,Zhao2018,Icaza2019,Ruggeri2019}. However, extracting the linear RSD signal from galaxy redshift surveys is non-trivial since the gravitational peculiar motions of galaxies are not fully linear and the RSD effect must be correctly modeled at nonlinear scales.

It has been shown that the growth rate can also be probed with cosmic voids. Indeed these underdense regions of matter, that account for about 80 per cent of the total volume of the observable Universe, are strongly affected by the growth of large-scale structure. Specifically, galaxies close to the edge of a void tend to be pushed away from the void centre, being attracted to the surrounding structure under the influence of gravity~\citep{Dubinski1993,Padilla2005}. These RSD introduce an anisotropy to the void-galaxy cross-correlation~\citep{Paz2013,Cai2016,Achitouv2017,Hamaus2015, Nadathur2019b} sensitive to the linear growth rate of structure. Recent measurements of the growth rate using voids have been performed on  BOSS~\citep{Hamaus2016,Hamaus2017,Nadathur2019,Achitouv2019, Hamaus2020}, 6dFGS~\citep{Achitouv2017} and VIPERS~\citep{Hawken2017}. 
Constraining the linear growth rate of structure using the RSD patterns around voids rather than on galaxies has several uses. Firstly, it is expected that, unlike the galaxy auto-correlation function,  which is quadratic in the density of galaxies, void-galaxy cross-correlation merely depends on galaxy density linearly, with reduced non-linear dynamics~\citep{Hamaus2014,Nadathur2019b}.  Secondly, the study of RSD around voids presents the opportunity to measure the growth of density perturbations in low-density regions. The comparison with the results from galaxy clustering in overdense regions is an attractive test for departures from Einstein gravity. 

Since the proof of the existence of voids in the distribution of galaxies~\citep{Gregory1978,Joeveer1978}, interest in using voids for cosmology has never ceased to grow~\citep{Lavaux2012}. As voids are nearly devoid of matter, they have proved to be very promising objects for exploring the imprint of possible modifications of GR such as f(R) gravity or extended gravity theories~\citep{Hui2009,Clampitt2013,Achitouv2016,Cai2015, Zivick2015,Voivodic2017,Cautun2018,Falck2018,Paillas2019,Perico2019} or the dark energy equation of state~\citep{Bos2012,Pisani2015}. Voids are also powerful probes to test the non-Gaussian nature of the primordial perturbation field~\citep{Kamionkowski2009}, to constrain the mass of neutrinos~\citep{Massara2015,Kreisch2019} or to investigate alternative dark matter scenario like warm dark matter~\citep{Yang2015}. 

In this work we perform an RSD analysis around cosmic voids using data samples from the extended Baryon Oscillation Spectroscopic Survey~\citep[eBOSS, ][]{Dawson2016} Data Release 16~\citep[DR16, ][]{DR16} of the Sloan Digital Sky Survey IV~\citep{Blanton2017}. eBOSS conducted a 5-year observation program, surveying the large scale structure of the Universe over a redshift range from 0.6 to 3.5. The eBOSS data samples we study are Luminous Red Galaxies (LRG), Emission Line Galaxies (ELG) and quasars (QSO). The construction of data catalogues is described in \citet[]{Ross2020,Lyke2020}, while mock catalogues are described in \citet[]{Zhao20}. The final eBOSS measurements of Baryon Acoustic Oscillation (BAO) and RSD in the clustering samples have been performed for LRG~\citep{Bautista20, Gilmarin20}, ELG~\citep{Raichoor2020,Tamone20,DeMattia20} and QSO~\citep{Hou20,Neveux20}. At the highest redshifts ($z>2.1$), the coordinated release of final eBOSS measurements includes measurements of BAO in the Ly-$\alpha$ forest~\citep{duMasdesBourboux20}. The multi-tracer analyses to measure BAO and RSD using LRG and ELG samples are presented in \citet[]{Wang20}. The cosmological interpretation of these results in combination with the final BOSS results and other probes is found in \citet{eBOSScosmo}. 

Prior to the final DR16 analysis, the signature of RSD around voids was already performed using the first two years of data from Data Release 14 (DR14) in \citet[]{Hawken2020}. Using DR16, we have six times more voids in the LRG and QSO samples compared to DR14, and we have for the first time a void catalogue derived from the ELG sample.

The paper is organized as follows. Section~\ref{sec:data} describes the DR16 galaxy samples and synthetic mock catalogues used in this analysis. Section~\ref{sec:void} presents the void finding routine applied to the aforementioned samples and the selection criteria applied to voids. In section~\ref{sec:analysis} we present the linear RSD model used to estimate the growth rate of structure in the DR16 sample, we describe its application on mocks and we evaluate systematic errors from different sources. In section~\ref{sec:results} we present the final constraints on the growth rate of structure using voids and finally conclude in section~\ref{sec:discuss}.

\section{Dataset}
\label{sec:data}
This study is part of a coordinated release of the final eBOSS measurements from the final release from SDSS-IV, DR16 \citep{DR16}. In this section, we describe the eBOSS DR16 datasets (Section~\ref{sec:data_dr16}) and present the synthetic mock catalogues that mimic the properties of the eBOSS data and that are used to compute the covariance and estimate systematic errors (Section~\ref{sec:data_mock}).

\subsection{Overview of the eBOSS survey}
Starting in 2014 with the fourth phase of the Sloan Digital Sky Survey program ~\citep[SDSS-IV;][]{Blanton2017}, the eBOSS survey~\citep{Dawson2016} was the successor of BOSS~\citep{Dawson2013}. The eBOSS targets were primarily observed using the BOSS double-armed spectrographs~\citep{Smee2013} on the 2.5-meter Sloan Telescope~\citep{Gunn2006}. 
A particular feature of the eBOSS survey is the use of four tracers of matter: LRGs in the redshift range $0.6<z<1.0$, ELGs in the redshift range $0.6<z<1.1$, QSOs used as direct tracers of the matter field in $0.8<z<2.2$, and higher redshift quasars ($z>2.1$) used for Ly$\alpha$ forest. The latter are excluded from the analysis presented here.

\subsection{DR16 data samples}
\label{sec:data_dr16}
The target selection of both LRG and QSO samples was conducted with the SDSS imaging photometry, a detailed description of these catalogues is given in the companion paper \citet{Ross2020}. The ELG target selection was done using the DECaLS part of the DESI Legacy Imaging Surveys\footnote{\url{http://legacysurvey.org/}} \citep{Dey2019} and the creation of the ELG catalogue is presented in the companion paper \citet{Raichoor2020}.
In this section we give a brief introduction to the data samples used in our analysis.

\subsubsection{The LRG sample}
The LRG sample was selected from the optical SDSS DR13 photometry~\citep{DR13} with additional publicly available infrared data from the WISE satellite~\citep{Wright2010}.  The final LRG selection is described in~\citet{Prakash2016}, for which colour cuts were applied to provide a sample with redshifts between $0.6<z<1.0$. The statistics for the eBOSS LRG sample are presented in \citet[table 4]{Ross2020}, with a total of 174,816 LRG over a footprint of 4,242 deg$^2$.

Following galaxy clustering analyses on the LRG sample in Fourier space~\citep{Gilmarin20} and configuration space~\citep{Bautista20}, we combine eBOSS LRGs with BOSS CMASS galaxies with $z>0.6$. The combined LRG+CMASS catalogue contains 377,458 galaxies with $0.6 < z < 1.0$ over a total footprint of 9,493 deg$^2$. All eBOSS LRGs are assumed to be within the CMASS footprint.

\subsubsection{The ELG sample}
ELGs are star-forming galaxies with strong emission lines, targeted as \oii doublet emitter at ($\lambda$3727, $\lambda$3729 \AA) for eBOSS. ELGs are primary targets in future spectroscopic surveys such as DESI~\citep[\oii emitter;][]{desi2016a,desi2016b} and Euclid~\citep[\halpha    \;emitter;][]{Amendola2018}.
The ELG selection performed in the DECaLS program~\citep{Dey2019} for eBOSS is described in \citet{Raichoor2017}. 
The building of the ELG catalogues for eBOSS DR16 is fully detailed in \citet{Raichoor2020}. This catalogue contains 173,736 ELGs between $0.6<z<1.1$ over a footprint of 1170 deg$^2$.

\subsubsection{The QSO sample}
The QSO sample covers a wide redshift range, bridging the gap between the CMASS galaxies at $z<0.7$ and the high redshift quasars at $z>2.2$ that probe the Ly$\alpha$ forest in the BOSS survey~\citep{Dawson2013}. The CORE QSO target selection is described in \cite{Myers2015}, using both optical imaging data from SDSS and mid-infrared data from the WISE survey~\citep{Wright2010}.  
The DR16 QSO catalogue is presented in \citet{Lyke2020} while the QSO clustering catalogue that we use is described in \citet{Ross2020}. 
The number of eBOSS QSOs is 343,708 covering a sky area of 4,808 deg$^2$ \citep[see][table 3]{Ross2020}, and spanning the redshift range $0.8<z<2.2$.

\subsubsection{Random catalogues}
For each of the above tracers, random catalogues are generated matching the angular and radial distribution of the data samples, but without any intrinsic clustering structure. The detailed description of the catalogue creation is given in \citet{Ross2020} and \citet{Raichoor2020} for the LRG and QSO samples and for the ELG sample, respectively.
The number density in random catalogues is at least 40 times larger than that of the data, in order to minimize shot noise. 

\subsubsection{Weights}

As galaxy redshift estimation depends on the observation conditions, weights are calculated to correct for possible systematic effects. These weights are used for creating void catalogues and for counting pairs when estimating the correlation function. They are briefly described here.

A few percent of targets are not observed due to fiber collisions. This happens when two or more galaxies are within 62$\arcsec$ and only one has an assigned fiber. The applied correction is to up-weight all objects in the same group by the close-pair weight $w_{\textsf{cp}}=N_{\textsf{targ}}/N_{\textsf{spec}}$, where $N_{\textsf{targ}}$ is the number of targets in the given group and $N_{\textsf{spec}}$ the number with spectroscopic observation. 
A similar weight $w_{\textsf{noz}}$ is defined for galaxies with no reliable redshift. The correction for redshift failure is based on the spectrograph signal-to-noise and the fiber ID. 
Similarly, to account for imaging systematics that generate spurious fluctuations in target selection, a weighting $w_{\textsf{sys}}$ is applied to each galaxy. 
Since the radial distribution of the tracers is not uniform but follows a radial mean density dependence n(z), an FKP weight is applied to objects in order to minimize the variance for clustering measurements, defined as~\citep{Feldman1994}:
\begin{eqnarray}
w_{\textsf{FKP}} = 1/[1+n(z)P_0],
\end{eqnarray}
where $P_0$ is the typical power spectrum value at the scale of BAO. For the different eBOSS tracers:
\begin{align}
P_{0, {\rm LRG}} &= 10000 \,h^{-3}\,{\rm Mpc}^3 , \\
P_{0, {\rm ELG}} &= 4000 \,h^{-3}\,{\rm Mpc}^3 , \\
P_{0, {\rm QSO}} &= 6000 \,h^{-3}\,{\rm Mpc}^3 .
\end{align}
The final weight for each galaxy can then be written as:
\begin{eqnarray}
\label{eq:weights}
w = w_{\textsf{noz}} \times w_{\textsf{cp}} \times w_{\textsf{syst}} \times  w_{\textsf{FKP}} .
\end{eqnarray}
This weighting scheme is the same for the data catalogue and the random catalogue.

\subsection{Mock catalogues}
\label{sec:data_mock}
In order to compute the covariance matrix and investigate systematic effects, we use synthetic mocks that mimic the data samples. 

\subsubsection{\EZ{}}
\EZ{} are fast generated mocks that encode effective structure formation and tracer bias models. They take into account radial distributions, veto masks and survey footprints as well as observational systematic effects. Mocks are used to compute the covariance matrix and to validate the analysis pipeline. 

\EZ{} are based on the Zel'dovitch approximation to generate a dark matter field at a given redshift~\citep{EZmocks}. The creation of mock catalogues for the LSS eBOSS tracers is extensively presented in \citet{Zhao20}. \EZ{} consist of a set of 1,000 realizations of light-cone mock catalogues for each type of tracers. For each of the \EZ{} realization is associated a random catalogue, as required for the normalization of clustering measurement and to fully simulate the dependence of random catalogues in observed data. The fiducial cosmological model used for constructing the \EZ{} is flat $\Lambda$CDM with: 
\begin{eqnarray}
\begin{split}
\Omega_{m} = 0.307, \,\, \Omega_{b} = 0.0482, \,\, h = 0.678,\\
\sigma_{8} = 0.8225, \,\, n_{s} = 0.96.
\end{split}
\end{eqnarray}
which are the best-fit values from the Planck 2013 results~\citep{Planck13}.

\subsubsection{\NSeries{} mocks}
\NSeries{} mocks are full N-body simulation populated with a single Halo Occupation Distribution (HOD) model. These mocks, which reproduce the BOSS CMASS LRG sample at the effective redshift $z=0.56$, are very useful to test model accuracy in the non-linear regime. A total of 7 independent periodic boxes projected through 12 different orientations for each box gives 84 pseudo-independent realizations for an effective volume of $84\times (2.6\;$\hgpc{})$^3$.

The underlying cosmology for \NSeries{} mocks is:
\begin{eqnarray}
\begin{split}
\Omega_{m} = 0.286, \,\, \Omega_{b} = 0.0470, \,\, h = 0.700,\\
\sigma_{8} = 0.82, \,\, n_{s} = 0.96.
\end{split}
\end{eqnarray}

\subsubsection{\OR{} mocks}
\OR{} mocks were created in the framework of the eBOSS mock challenge whose purpose was to provide N-body based mocks to study eventual systematic effects of the HOD models on standard galaxy clustering measurements. Those mocks are based on the N-body \OR{} simulation \citep{OuterRim,Heitmann_2019a,Heitmann2019} of $10,240^3$ particles in a $(3\;$\hgpc$)^3$ volume and built from snapshots of the simulation.

The underlying cosmology for \OR{} simulation is close to the best-fitting model from WMAP-7~\citep{Komatsu2011}:
\begin{eqnarray}
\begin{split}
\Omega_{m} = 0.2648, \,\, \Omega_{b} = 0.0448, \,\, h = 0.71,\\
\sigma_{8} = 0.8, \,\, n_{s} = 0.963.
\end{split}
\end{eqnarray}

\paragraph{\OR{} ELG mocks}
\OR{} ELGs are built from a single snapshot at $z=0.865$, close to that of the DR16 ELG sample. Six sets of mocks were produced, each with a different HOD model. The detailed description of the mock construction and HOD models can be found in \citet{Alam20}. In this paper, we use one blind mock of the ELG mock challenge with a galaxy number density similar to that of the data and populated with the HMQ3 (HighMassQuenched-3) HOD model. This mock contains 30 pseudo-independent realizations with periodic boundary conditions.

\paragraph{\OR{} QSO mocks}
\OR{} QSOs are built from a snapshot at $z=1.433$. From this snapshot, 20 sets of mocks were created and populated with 20 different HOD models. In order to include the effect of quasars redshift uncertainties, an additional redshift smearing was added to mocks, providing 4 variations of the same mock with a redshift smearing of varying intensity. The detailed description of the mock construction, HOD modelling and redshift smearing along with their impact on standard clustering measurements are described in \citet{smith20}. We use a 'non-blind' mock populated with the HOD10 model with a prescription of a realistic redshift smearing case. It contains 100 pseudo-independent realizations with a tracer density comparable to that of the QSO sample.

\section{Void catalogues}
\label{sec:void}
In this section we present the construction of void catalogues from the data and \EZ{} in eBOSS DR16 samples. We describe the main steps of the void finding algorithm (Section~\ref{sec:void_finding}) and present the selection cuts applied to remove voids too close to the survey edge  (Section~\ref{sec:void_selection}). We then present statistics of final void catalogues and compare basic properties of voids between data and \EZ{} (Section~\ref{sec:void_stat}). 

\subsection{Void finding algorithm}
\label{sec:void_finding}
\textsc{Revolver} \footnote{\url{http://github.com/seshnadathur/Revolver}}  \citep{Nadathur2019} 
is  a multi-purpose algorithm that applies both reconstruction and void-finding on a given galaxy or simulated data sample. We make use of the void finding part of the algorithm only, without applying prior reconstruction. 

Prior to any void finding, the galaxy positions are transformed to comoving space in \hmpc{} assuming a flat $\Lambda$CDM cosmology with $\Omega_m = 0.31$. 

The void finding part of \textsc{Revolver} is comprised of a python wrapper around the ZOBOV algorithm \cite{Neyrinck2008}.
The ZOBOV algorithm performs a Voronoi Tessellation Field Estimation (hereafter VTFE) on the discrete sample of tracers : each tracer is assigned a cell which encompasses all the nearest points to the considered tracer. This process allows an estimation of a local volume associated with a given tracer. By definition of the VTFE, the inverse of the estimated volume provides a measure of the local density within each cell.
Local density minima in the tessellation field are then identified and adjacent low density galaxies are merged in order to form zones of minimal density without density threshold. This process is re-iterated for the zones, allowing us to identify low density regions throughout the survey footprint : these regions are called voids. 

\textsc{Revolver}  applies a rescaling to the volumes estimated through the VTFE in order to take into account both the selection function and weights correcting for systematics in the survey, with the following association :  $V^{\mathrm{res}}_j =  V_j * w_{z} / w_{\mathrm{tot}}$ where $V_{j}$ is the volume of the Voronoi cell enclosing the galaxy $j$, $w_z$ is the weight arising from the selection function estimated in the void finder and $w_{tot}$ is the combined systematic weights defined in Eq.~\ref{eq:weights} without the $w_{\mathrm{FKP}}$ contribution. 

In order to practice a consistent tessellation of the density field and avoid leakage at the boundary of the survey both in redshift and footprint, buffer particles are positioned along the survey boundaries with a density of  $100 \; \bar{n_g}$.  The galaxies are checked for any proximity to these particles and are flagged not to be trusted in case of adjacency.  
Underdense zones processed from ZOBOV are then flagged as edge if considered too close to the boundary because of the higher probability of their volume to be ill-defined.  

In the post-processing part, zones are separated if needed  in order to obtain the smallest entity corresponding to an under-density.  It differs from other ZOBOV-based void finders \citep{Sutter2014} VIDE in this sense, because it does not try to probe the void hierarchy, finding only what would be called child void. 
This should not affect the making of our samples, as no prior void samples made from available galaxy datasets using ZOBOV-like algorithm have managed to be sensitive to the void hierarchy. 
The centre of such a void is then defined as the volume-weighted barycentre of the galaxies defining the void.   
An effective radius is estimated from the total volume of the voids taken as that of a sphere : 

\begin{eqnarray}
r_v = \bigg(\frac{3}{4\pi}\sum_j V_j\bigg)^{1/3}.
\end{eqnarray}
where $V_j$ is the volume associated with the Voronoi cell of the galaxy $j$ used to define the void and its barycentre. 
All properties pertaining to the voids use the non-rescaled voronoi volume $V_j$ to compute the properties, while the rescaled density $\rho^{\mathrm{res}} = 1/V^{\mathrm{res}} $ is used as a weight to take into account the systematic effects in the void properties definition.

\subsection{Selection cuts}
\label{sec:void_selection}
A drawback in the void finding procedure is the effect of the proximity of buffer particles positioned at the boundary of the survey. Although these particles prevent us from finding voids in the vetoed portions of the survey, their presence causes an increase in spurious voids that cannot be distinguished from the "true" under-densities in the density field.  As a result, we apply three specific selection cuts to keep only those voids that we consider to be reliable in our final samples.

\begin{description}

\item \texttt{Npart cut}: 
Any voids defined by less than five galaxies are excluded from the void catalogue, as they are considered to be poorly defined voids.

\item \texttt{Edge Flag cut}: 
Any voids with a non null \texttt{Edge flag} value are discarded from the void catalogue as their volume and properties are inclined to be ill-defined through proximity to buffer particles.

\item \texttt{NearestEdge cut}: 
Any voids too close to the redshift boundaries are also removed. Since many buffer particles are created for the needs of the void finding, their presence causes an increase of the number of voids near the redshift boundaries. To mitigate this effect, we discard all voids for which the position of the void centre added to the effective radius $r_v$ of the void or added to the distance of the farthest galaxy belonging to the void exceeds the distance of the nearest limit in redshift.
\end{description}

\subsection{Final void catalogues}
\label{sec:void_stat}

\subsubsection{Statistics}
The summary statistics of void catalogues for each sample of synthetic \EZ{} are presented in Table~\ref{table:stats}. The number of voids before and after selection cuts are averaged over the 1,000 realizations of each tracer. It is mostly the LRG+CMASS sample that suffer severe cuts with the set defined in section~\ref{sec:void_selection}.

Table~\ref{table:stats} also shows the summary statistics of void catalogues for the three eBOSS DR16 data samples. These quantities are subject to small fluctuations due to the inherent procedure of the void finder. Indeed the number of buffer particles that are added to galaxy or quasar catalogues to prevent the algorithm from finding voids outside the survey boundary has an effect on the void finding process. Since these particles are randomly positioned along the boundaries of the veto mask, the calculation of the volume of the Voronoi cells may be slightly modified from one realization of void finding to another, which leads to some fluctuations in terms of void statistics, the resulting catalogues being slightly different. 
To circumvent this problem, we apply the \textsc{Revolver} algorithm 1,000 times on each data catalogue. The analysis described in section~\ref{sec:analysis} will be systematically applied to all of these 1000 catalogues, for each data sample, unless otherwise stated. 
Statistics presented in Table~\ref{table:stats} are given in terms of means over the 1000 void data catalogues generated with \textsc{Revolver}. The related systematic uncertainty is estimated in section \ref{sec:ana_syst}.

\begin{table*}
    \centering
    \caption{Statistics of void catalogues identified in \EZ{} catalogues and eBOSS DR16 LSS catalogues. The quantity $N_{\textsf{g}}$ is the number of galaxies or quasars, $N_{\textsf{v}}$ and $N_{\textsf{v,cut}}$ are the number of voids before and after selection cuts as described in section~\ref{sec:void_selection}, respectively. The numbers $N_{\textsf{v}}$ and $N_{\textsf{v,cut}}$ are averages over the 1000 realizations and the error on these mean values are typically of the order of $\pm 2$. The quantity $z_{\textsf{eff}}$ is the effective redshift of the void catalogues after selection cuts. } 
    \begin{tabular}{lccccccc}
    \hline 
    Sample & $N_{\textsf{g}}$ & $N_{\textsf{v}}$ & $N_{\textsf{v,cut}}$  &  z range & $z_{\textsf{eff}}$ & $r_\textsf{max}$ & Area ($\mathrm{deg}^2$)  \\
    \hline \hline
    \textbf{\EZ{}} \\
    LRG+CMASS & 380,190 & 4,283  & 2,832  &  $0.6<z<1.0$ & 0.740 & 3.52 & 9,493 \\  
    ELG       & 173,736 & 2,209  & 1,895  &  $0.6<z<1.1$ & 0.847 & 3.60 & 1,170 \\
    QSO       & 343,700 & 5,449  & 4,321  &  $0.8<z<2.2$ & 1.478 & 3.52 & 4,808 \\
    \hline
    \textbf{Data sample}  \\
    LRG+CMASS & 377,458 & 4,228 & 2,814 &  $0.6<z<1.0$ & 0.740 & 3.52 & 9,493 \\
    ELG       & 173,736 & 2,097 & 1,801 &  $0.6<z<1.1$ & 0.847 & 3.60 & 1,170 \\
    QSO       & 343,708 & 5,451 & 4,347 &  $0.8<z<2.2$ & 1.478 & 3.52 & 4,808 \\
    \hline \hline
    \end{tabular}
    \label{table:stats}
\end{table*}

In order to define the effective redshift of the void sample, we perform the following weighted void-galaxy pair-count,

\begin{eqnarray}
\label{eq:eff_z}
z_{\textsf{eff}} = \frac { \sum_{ij} w_i (z_i + Z_{j})/2} 
{ \sum_{i} w_i },
\end{eqnarray}
where $z_i$ is the redshift of the $i^{th}$ galaxy, $Z_j$ the redshift of the centre of the $j^{th}$ void, and $w_i$ the total weight of the $i^{th}$ galaxy, as given by Eq~\ref{eq:weights}. The computation is made over all void-galaxy pairs used for the correlation function in the range $[0 - r_{\textsf{max}}[$, where $r_{\textsf{max}}$ corresponds to the maximal radial separation between the void centre and the galaxy, rescaled by the void radius $r_v$ of the considered void.
The subsequent effective redshifts and their corresponding $r_{\textsf{max}}$ are given in Table~\ref{table:stats} for each eBOSS sample.

\subsubsection{Redshift distribution}
\begin{figure}
    \includegraphics[width=\columnwidth]{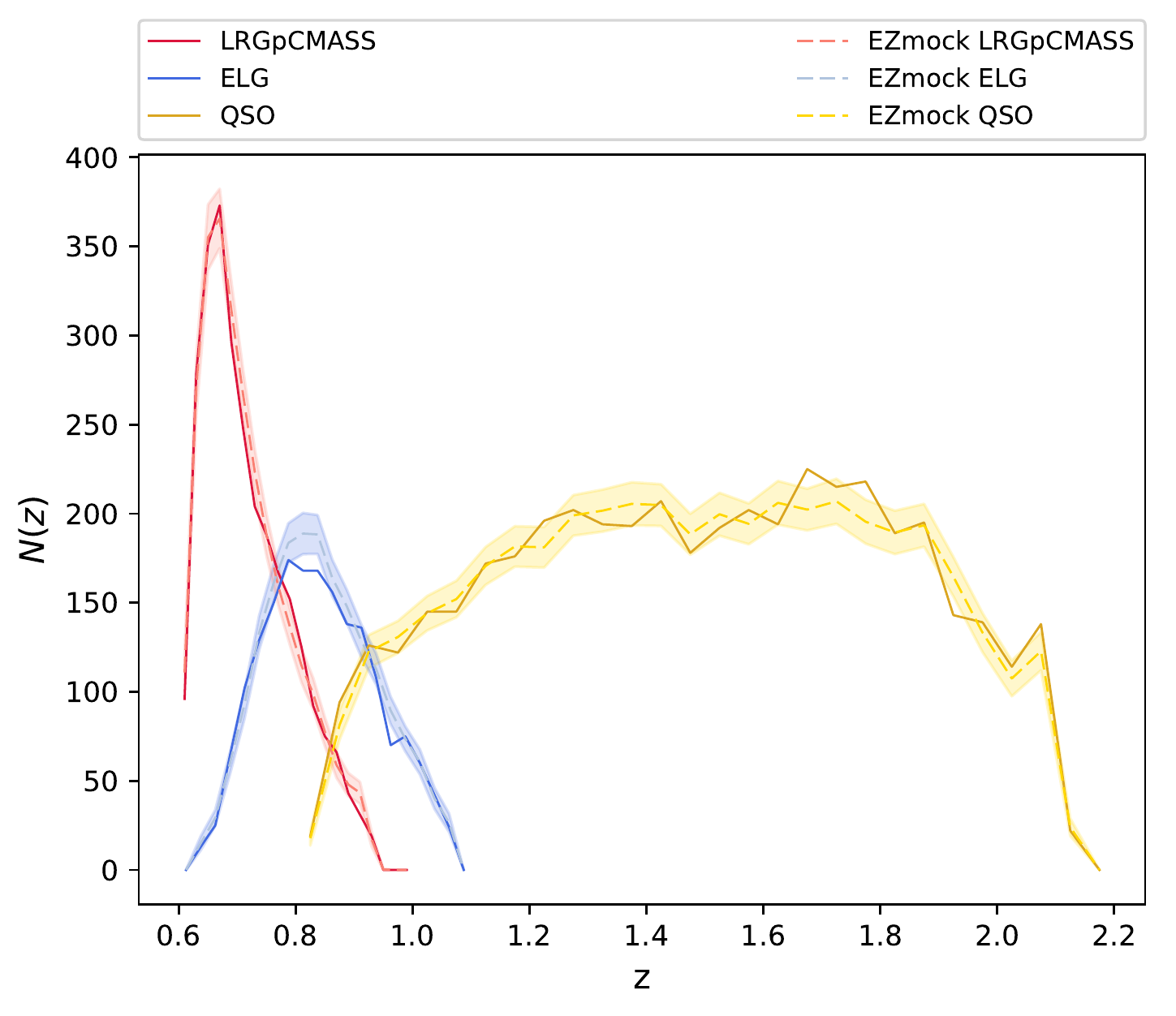}
    \caption{\label{fig:redshift} Redshift distribution of voids after selection cuts for LRG+CMASS samples (red lines), ELG samples (blue lines) and QSO samples (yellow lines). The solid and dashed lines correspond to the data and the mean of the 1000 realizations of the \EZ{}, respectively. The shaded areas indicate the $1\sigma$ regions evaluated from 1000 mock realizations.}
\end{figure}

Figure~\ref{fig:redshift} shows the redshift distribution for the three tracer populations in eBOSS. The \EZ{} (dashed lines) averaged over 1000 realizations within $1\sigma$ dispersion (shaded areas) are compared to the data samples (solid lines) for the LRG+CMASS, ELG and QSO samples. There is a good agreement between voids found in mock catalogues and those from data. The asymmetric distribution of LRG+CMASS voids with an excess towards low redshifts results from the population of CMASS galaxies added to the eBOSS LRGs.

\subsubsection{Abundances} 

\begin{figure}
    \includegraphics[width=\columnwidth]{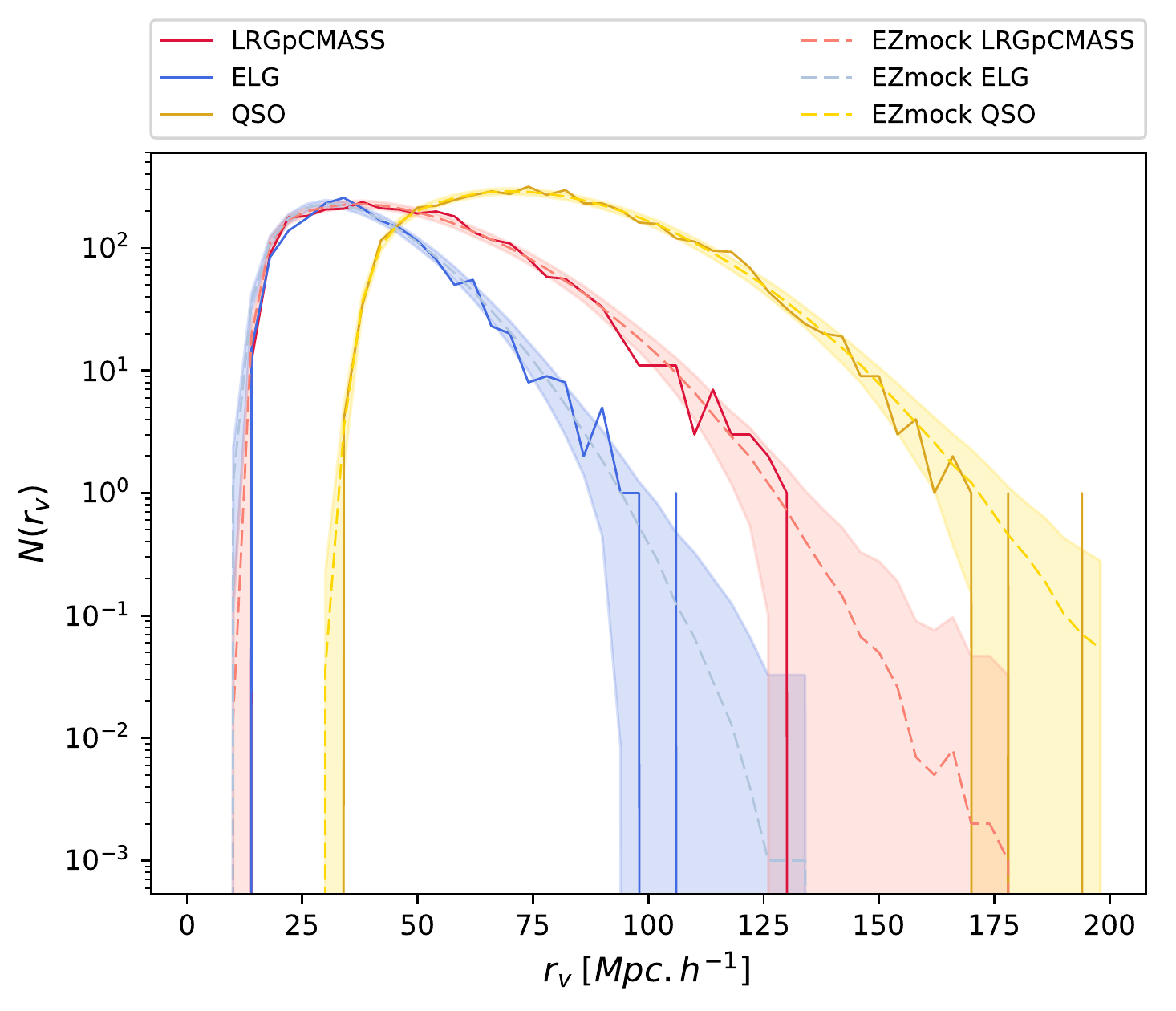}
    \caption{\label{fig:abundance} Number of voids after selection cuts as a function of their radius $r_v$ for LRG+CMASS samples (red lines), ELG samples (blue lines), and QSO samples (yellow lines). The solid and dashed lines correspond to the data and the mean over the 1000 realizations of the \EZ{}, respectively. The shaded areas indicate the $1\sigma$ regions evaluated from 1000 mock realizations.}
\end{figure}

Figure~\ref{fig:abundance} displays the distribution of the number of voids as a function of their radius $r_v$, for the three types of eBOSS tracers. Voids are, on average, larger in the quasar sample than in the galaxy samples, with sizes up to 175\;\hmpc{}, compared to 125\;\hmpc{} and 100\;\hmpc{} for LRGs and ELGs, respectively. Several authors have underlined that the number counts of cosmic voids detected in galaxy surveys may depend on the tracer bias~\citep{Pollina2019} and on the sparsity of the survey
~\citep{Jennings2013, Sutter2014b}. Indeed, as the algorithm tessellates the discrete distribution of galaxies, we expect voids to be larger as the density of the survey decreases. 

Although void abundance can be useful to provide constraints on dark energy or modified gravity models~\citep{Pisani2015,Voivodic2017,Verza2019}, we only use them here to make basic comparisons between the data and synthetic catalogues, in order to validate mocks for void analysis. Figure~\ref{fig:abundance} also shows the mean of the void count distribution over the 1,000 mocks of each sample of \EZ{}, while the $1\sigma$ dispersion is indicated by the shaded area. The comparison between the data (solid lines) and the synthetic \EZ{} (dashed lines) shows a good agreement for the LRG+CMASS, ELG and QSO samples in terms of void counts. 

\section{Methodology}
\label{sec:analysis}
In this section we describe the void-galaxy clustering estimation and the modelling of redshift space distortions (Section~\ref{sec:ana_cor}). We present the steps of the fitting procedure (Section~\ref{sec:ana_fit}). Then, once we have validated the clustering properties of the \EZ{} against the data (Section~\ref{sec:ana_comp}), we perform the fit procedure on the mocks in order to extract the cosmological information by measuring the distortion parameter $\beta$ (Section~\ref{sec:ana_mock}). This value is used as a reference value for systematic studies (Section~\ref{sec:ana_syst}).

\subsection{The void-galaxy cross-correlation function}
\label{sec:ana_cor}

\subsubsection{Cross-correlation function estimator}
The void-galaxy cross-correlation function $\xi^s(r,\mu)$ describes the density contrast around voids in redshift space, $ \delta(r) = \rho(r) / \bar{\rho}(r)  -1$, where \textit{r} is the void-galaxy separation distance normalized to the effective radius of the void $r_v$.

For extracting the void-galaxy clustering information we can either extend the Landy-Szalay estimator~\citep[LS; ][]{LandySzalay93} as: 
\begin{eqnarray}
\label{eq:LS}
\xi^{LS}(r,\mu) = \frac{D_vD_g-D_vR_g-D_gR_v+R_vR_g}{R_vR_g} ,
\end{eqnarray}
or use the Davis-Peebles estimator~\citep[DP; ][]{Davis1983}:
\begin{eqnarray}
\label{eq:PD}
\xi^{DP}(r,\mu) = \frac{D_vD_g}{D_vR_g} - 1,
\end{eqnarray}
where D refers to the data and R to the randoms, the subscript \textit{v} refers to the voids and the subscript \textit{g} to the galaxies, and each pair \textit{XY} refers to the number of void-galaxy pairs at a distance \textit{r} normalized to the radius $r_v$ of the void. 

Although the consensus estimator in galaxy clustering is usually the LS-estimator, the choice of the estimator is more tricky in the case of voids. 
Some authors adopt the LS-estimator to compute the void-galaxy cross-correlation function~\citep{Achitouv2019,Nadathur2019}. The production of realistic random void catalogues is highly non-trivial. Voids are extended objects that, following our definition, are also mutually exclusive. One possible method to produce a random catalogue of voids might be to run our void finder on the same random catalogue, with the same number density as our galaxy catalogue. However, it is not clear if this would produce a random void catalogue with the correct properties to use in eq.~\ref{eq:LS}. In addition,  \citet[]{Hamaus2017} point out that the contribution of the terms involving $R_v$ is negligible in the multipole terms of the void-galaxy cross-correlation function. We therefore choose to employ the DP-estimator as in our previous work~\citep{Hawken2020}.

\subsubsection{Linear redshift space distortions}
Due to redshift space distortions (RSD) resulting from peculiar velocities of galaxies around voids, the pattern of the voids is distorted, leading to an anisotropic cross-correlation function. The void-galaxy cross-correlation function as estimated from Eq.~\ref{eq:PD}.
 can therefore be decomposed in terms of multipole moments $\xi_\ell(r)$ on the basis of Legendre Polynomials $\mathcal{L}_{\ell}(\mu)$:

\begin{eqnarray}
\xi^{s}(r,\mu) = \sum_\ell \mathcal{L}_\ell(\mu) \xi_\ell(r),
\end{eqnarray}
where $\mu$ is the cosine of the angle between the separation vector direction $r$ and the line-of-sight, and $\xi_\ell(r)$ the multipole defined as:
\begin{eqnarray}
\xi_\ell(r) = (2\ell + 1) \int_0^1 \mathcal{L}_\ell(\mu) \xi(r,\mu) d\mu.
\end{eqnarray}
We note that all odd multipoles cancel out. 

In the case of voids, the modelling of the apparent distortions is remarkably well described by linear theory~\citep{Hamaus2015}. In this paper we consider the linear model of RSD as proposed by~\citet{Cai2016}, in which voids are considered stationary, leading to only monopole ($\ell = 0$) and quadrupole ($\ell = 2$) non null terms.

The two point correlation function thus reduces to:
\begin{eqnarray}
\label{eq:anisxivg}
\xi^s(r,\mu) = \mathcal{L}_0(\mu) \xi^s_0(r) +  \mathcal{L}_2(\mu) \xi^s_2(r) ,
\end{eqnarray}
with first order Legendre polynomials: 
\begin{eqnarray}
\label{eq:Lpoly}
\mathcal{L}_0(\mu) &=& 1 ,\\
\mathcal{L}_2(\mu) &=& \frac{3\mu^2 - 1}{2},
\end{eqnarray} 
and the resulting multipoles can be written as:
\begin{eqnarray}
\label{eqn:monopole}
\xi^s_0(r) &=&  (1+\frac{\beta}{3}) \; \xi(r),
\end{eqnarray} 
\begin{eqnarray}
\label{eqn:quadrupole}
\xi^s_2(r) &=& \frac{2\beta}{3} \; [ \xi(r) - \bar{\xi}(r) ] ,
\end{eqnarray} 
where $\beta$ is the linear redshift distortion parameter defined as $\beta = f/b$, with $f$ the linear growth rate of density perturbations and $b$ the linear galaxy bias, and
\begin{eqnarray}
\label{eqn:xibar}
\bar{\xi}(r) = \frac{3}{r^3}\int_0^r \xi(r')r'^2 {\rm d}r'.
\end{eqnarray}

By combining Eq.~\ref{eqn:monopole} and Eq.~\ref{eqn:quadrupole}, an estimate of the distortion parameter is given by~\citep{Cai2016}:
\begin{eqnarray}
G(\beta) & = & \frac {\xi^s_2(r)} {\xi^s_0(r) - \bar{\xi}^s_0(r)} \\
& = & \frac {2\beta} {3+\beta} .
\end{eqnarray}
In practice we will minimize the residual:
\begin{eqnarray}
\label{eqn:residuals}
  \epsilon(\beta) = \xi_2 -  ( \xi_0 - \bar{\xi}_0 ) \;   \frac {2 \beta} {3+\beta} .
\end{eqnarray}

This model is a first-order derivation of linear perturbation theory. It has been found to be effective in measuring the growth rate of structures in previous analyses~\citep{Hamaus2017,Achitouv2019,Hawken2020} and requires almost no knowledge of the true correlation function between void and galaxy, which has no theoretical formulation yet (except for fitting functions) nor specific modelling of peculiar velocities such as the Gaussian Streaming Model~\citep{Hamaus2015}. 

\subsection{The fitting procedure}
\label{sec:ana_fit}
The linear growth rate estimation is performed by means of $\chi^2$ minimisation, where the $\chi^2$ is defined as:
\begin{eqnarray}
\label{eqn:chi2}
\chi^2 = \epsilon^T \; \Psi \; \epsilon ,
\end{eqnarray}
where $\epsilon$ is the residual given by Eq.~\ref{eqn:residuals} and $\Psi$ is the precision matrix. An unbiased estimate of the precision matrix, which compensates for the bias present when inverting a noisy covariance matrix, is given by~\citep{Hartlap2007, Taylor2012} 
\begin{eqnarray}
\hat{\Psi}=\frac{N_s - N_b - 2}{N_s - 1} \hat{C}^{-1},
\end{eqnarray}
where $N_b$ is the number of bins and $N_s$ is the number of mocks used to estimate the covariance matrix $\hat{C}$. For a covariance matrix derived from $N_s=1000$ \EZ{} realizations, and with around 20 measurement bins for each sample, the correction factor is less than $2\%$ in our uncertainty estimates. 

The covariance matrix $\hat{C}$ is estimated for each tracer with their 1000 \EZ{} realizations presented in Sec~\ref{sec:data_mock}. The covariance is computed as follows: 
\begin{eqnarray}
\label{eq:cov}
C_{ij} = \frac{1}{N_s - 1}\sum_{k=1}^{N_s}(\epsilon_i^k - \langle \epsilon_i \rangle)(\epsilon_j^k - \langle \epsilon_j \rangle),
\end{eqnarray}
where $N_s$ is the number of independent mocks, $\epsilon^k_i$ is the residual of the mock $k$ in the bin $i$ and $\langle \epsilon_i \rangle$ is the mean value of $\epsilon^k_i$ in the bin $i$ such as:
\begin{eqnarray}
\label{eq:moy}
\langle \epsilon_i \rangle = \frac {1}{N_s} \sum_{k=1}^{N_s} \epsilon_i^k .
\end{eqnarray}



The best-fitting parameter is found by minimizing the $\chi^2$ using the MINUIT algorithm~\citep{MINUIT}. The uncertainty in the covariance matrix estimate is propagated in the fitted parameter errors following prescriptions described in \citet[]{Percival2014, Dodelson2013}.

\subsection{Comparing void clustering in data and mocks}
\label{sec:ana_comp}
\begin{figure*}
\begin{subfigure}{2\columnwidth}
    \caption{LRG+CMASS}
\includegraphics[width=1.\columnwidth, height=6cm]{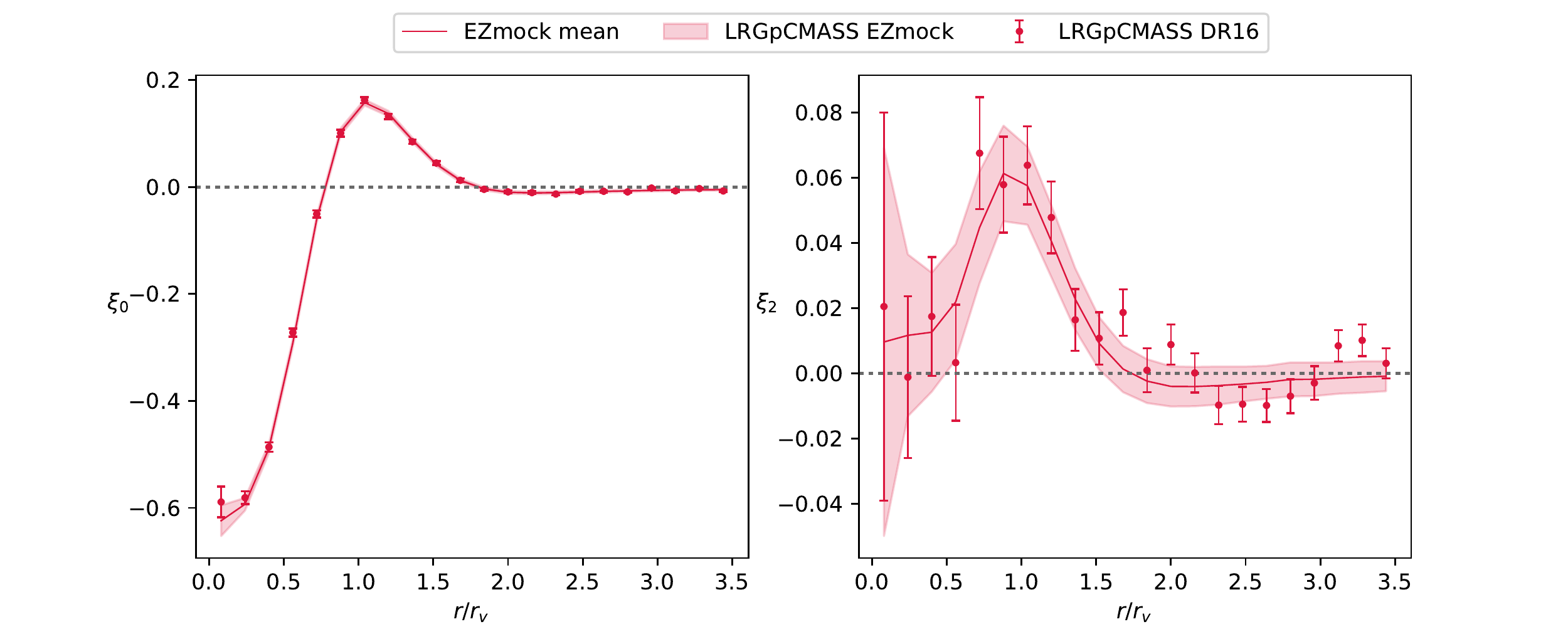}
    \label{fig:LRGcorr}
\end{subfigure}
\begin{subfigure}{2\columnwidth}
    \caption{ELG}
\includegraphics[width=1.\columnwidth, height=6cm]{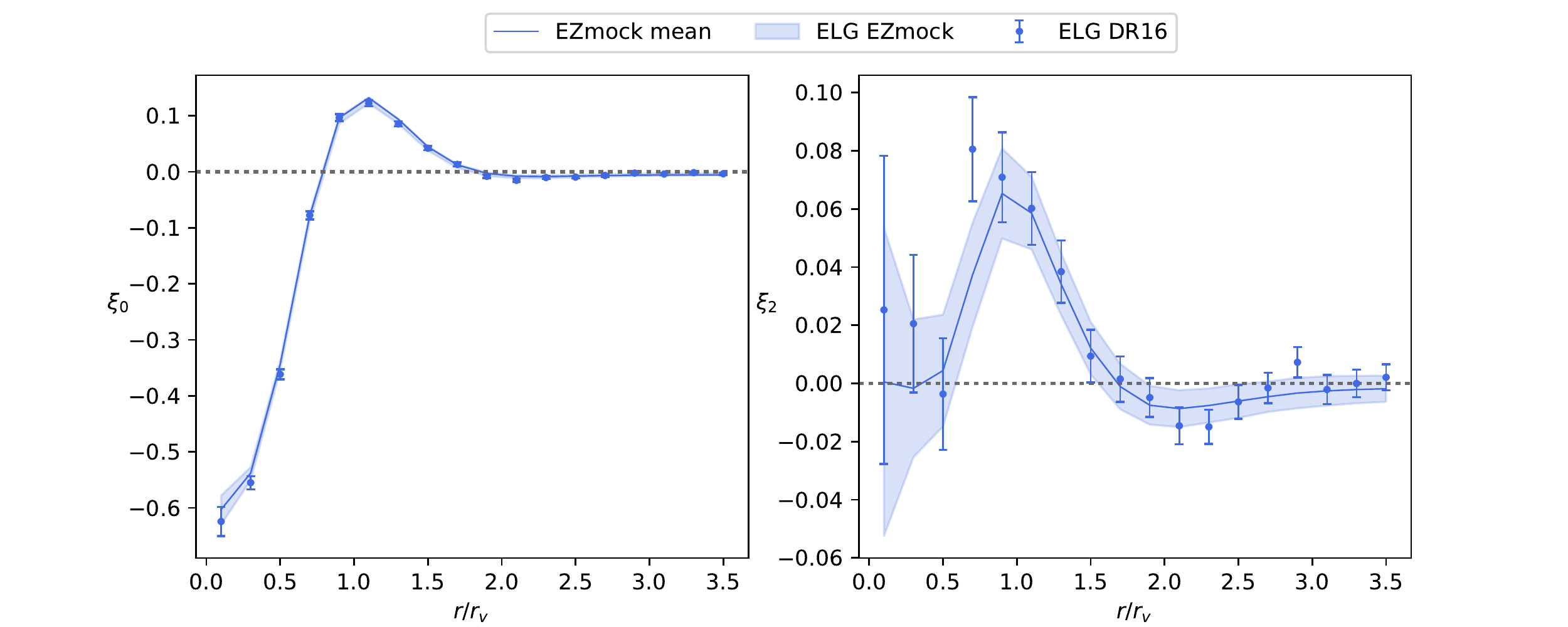}
    \label{fig:ELGcorr} 
\end{subfigure}
\begin{subfigure}{2\columnwidth}
    \caption{QSO}
    \label{fig:QSOcorr}
\includegraphics[width=1.\columnwidth, height=6cm]{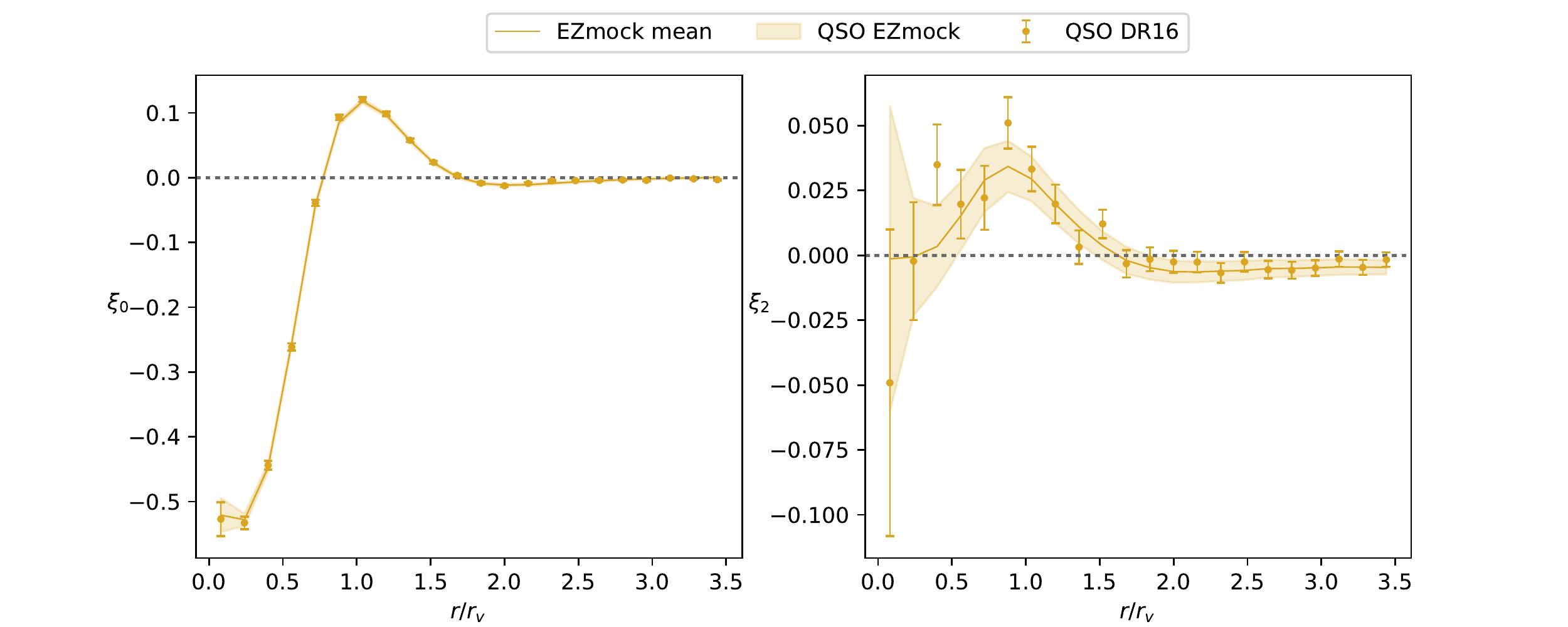}
\end{subfigure}
\caption{Multipoles of the DR16 void-galaxy cross-correlation functions of data compared to the mock catalogues. Left panels show the monopole component and right panels show the quadrupole component, as a function of the separation distance $r$ normalized to the effective void radius $r_v$. The LRG+CMASS, ELG and QSO DR16 samples are displayed in the top (a), middle (b) and bottom (c) panels, respectively, for the data (circle symbol) and the mean of 1000 \EZ{} realizations (solid line). The shaded region shows the standard deviation of the 1000 mock realizations, and error bars on data are the square-root of the diagonal elements of the covariance matrix.}
\label{fig:DR16_corr}
\end{figure*}

Figure~\ref{fig:DR16_corr} displays the void-galaxy cross-correlation function for one realization of the DR16 data samples and the mean of the 1000 \EZ{} realizations. The subpanels~\ref{fig:LRGcorr}, \ref{fig:ELGcorr} and \ref{fig:QSOcorr} show the LRG+CMASS, ELG and QSO samples, respectively. The left panels display the monopole $\xi_0$ and the right panels the quadrupole $\xi_2$ of the correlation function.

The monopole of the cross-correlation is indicative of the mass-density profile in voids~\citep{Hamaus2014}. It exhibits a deep under-dense core near the centre of the void at $r<0.5 r_v$ and an overdense compensation wall close to the edge of the void at $r=r_v$.  At sufficiently large distances from the void centre ($r>2r_v$), the density tends towards the mean background density. The shape of the density profile of voids was shown to be universal and can be parametrized by an empirical function~\citep{Hamaus2014b,Ricciardelli2014,Nadathur2015}. However, given the fitting parametrization in Eq.~\ref{eqn:residuals} where all quantities are measured from data, there is no need to assume a density profile.

The comparison of the void-galaxy cross-correlation function between the data and the average of the \EZ{} seems to match nicely for both the monopole and the quadrupole. This good agreement confirms that we can use \EZ{} to test our fitting procedure before applying it in a blinded way to our data.

\subsection{Fitting mock catalogues}
\label{sec:ana_mock}
In this section, we present tests on our distortion parameter fitting methodology applied to mocks. We will investigate the results
from the mean of the \EZ{} and perform an optimization of the procedure using these results. 

\begin{figure}
\begin{subfigure}{\columnwidth}
\caption{\EZ{} LRG+CMASS}
\includegraphics[width=\columnwidth, height=5cm]{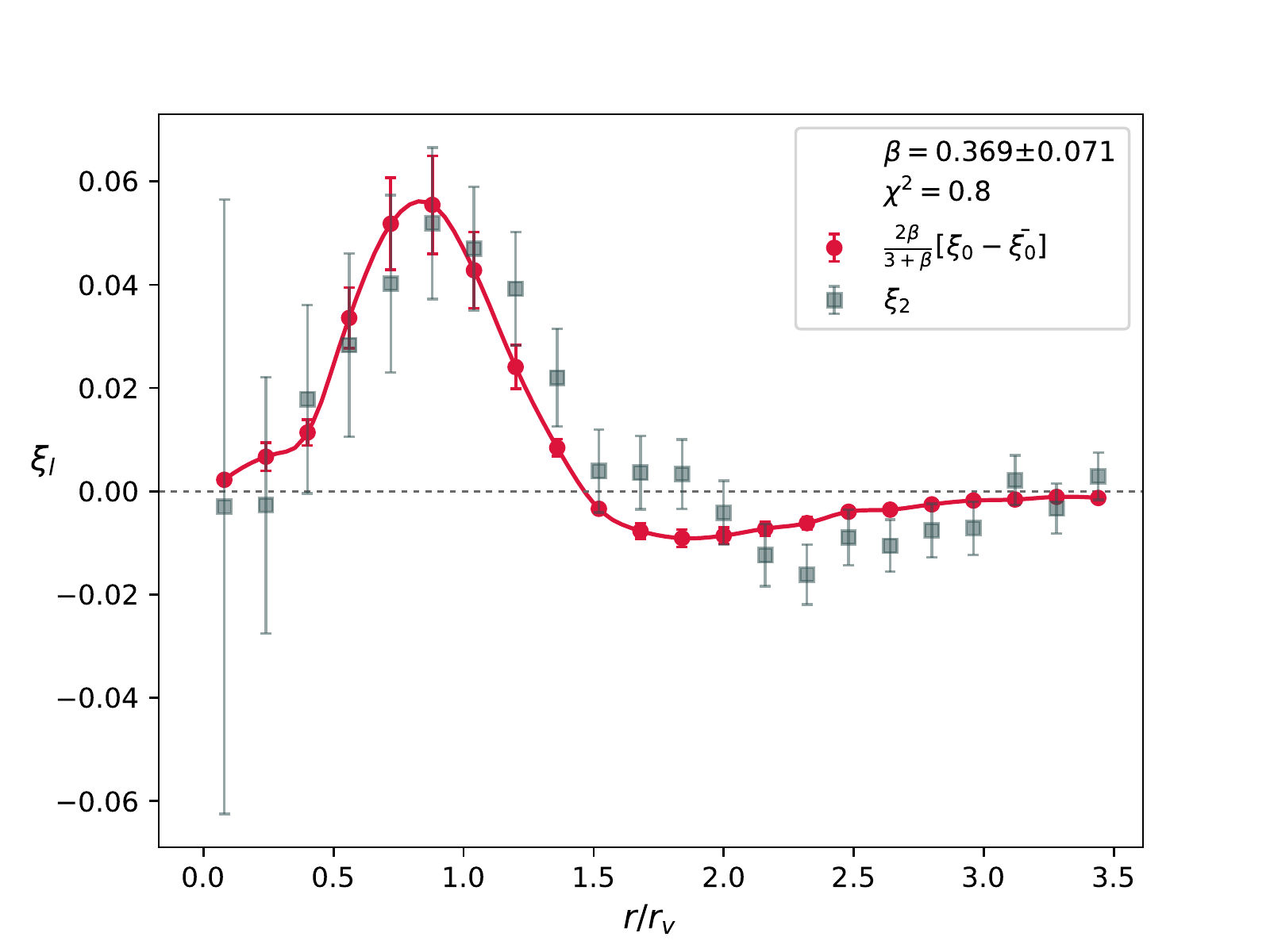}
\label{subfig:LRG_bf_mock} 
\end{subfigure}
\begin{subfigure}{\columnwidth}
\caption{\EZ{} ELG}
\includegraphics[width=\columnwidth, height=5cm]{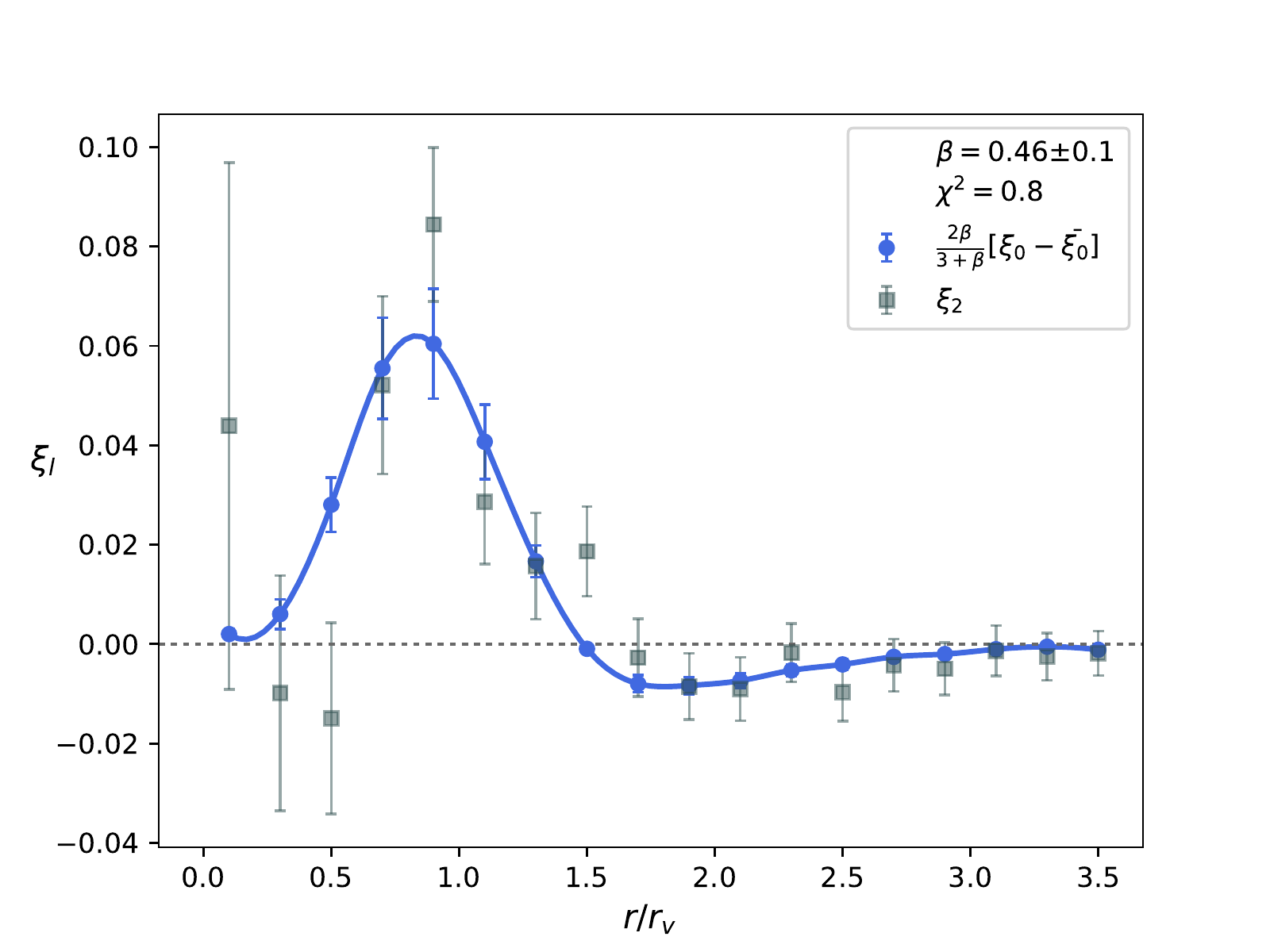}
\label{subfig:ELG_bf_mock}
\end{subfigure}
\begin{subfigure}{\columnwidth}
\caption{\EZ{} QSO}
\includegraphics[width=\columnwidth, height=5cm]{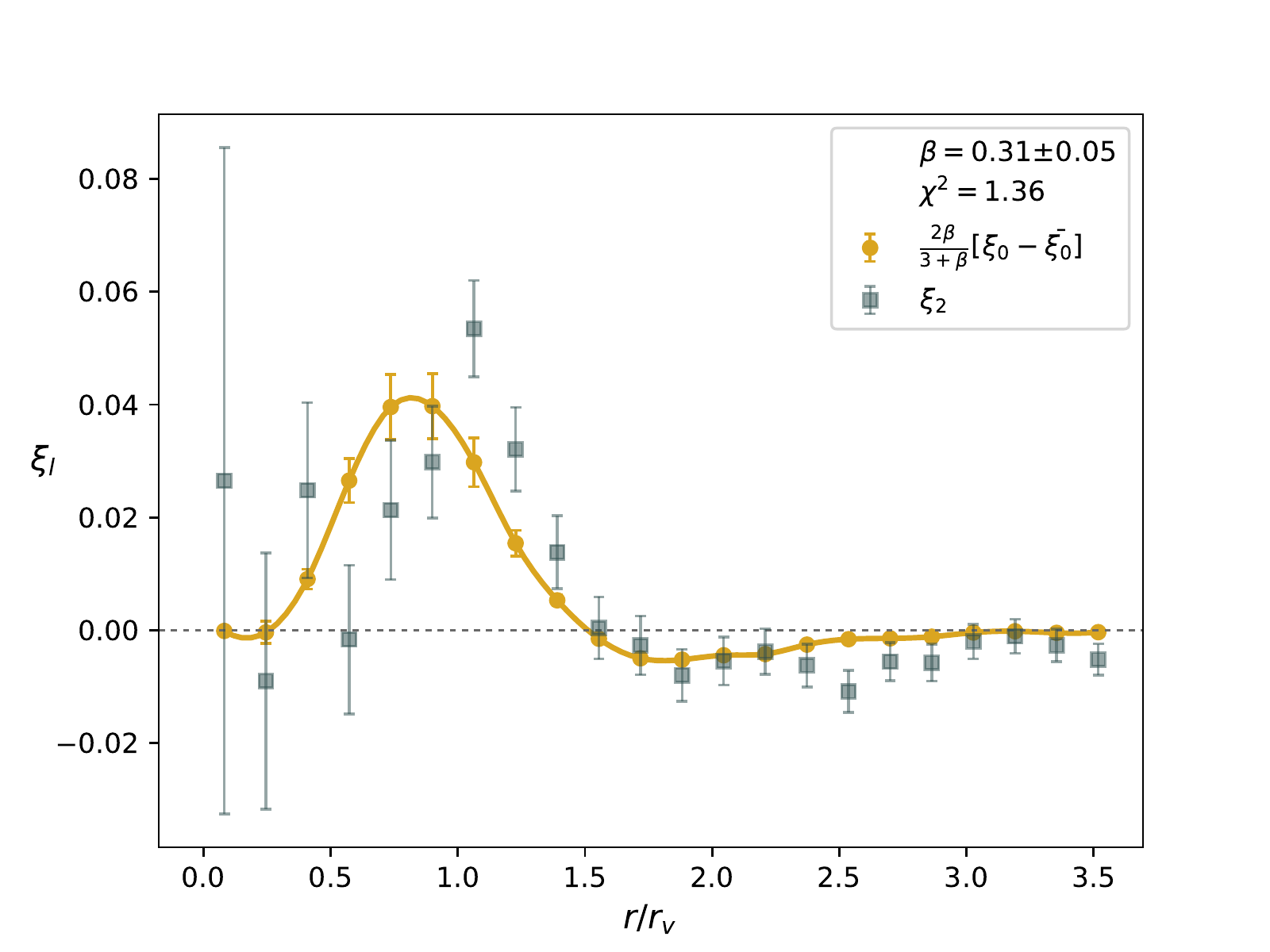}
\label{subfig:QSO_bf_mock} 
\end{subfigure}
\caption{Quadrupole ($\xi_2$) and the best-fit of the $ 2 \beta / (3 + \beta) (\xi_0-\bar{\xi_0})$ from one \EZ{} catalogue of the LRG+CMASS, ELG and QSO sample displayed in the top (a), middle (b) and bottom (c) panels, respectively. Error bars are the diagonal of the covariance matrix from the $N_s-1$ remaining mocks.
}
\label{fig:Bestfit_mock}
\end{figure}

Multipoles of the void-galaxy cross-correlation function are computed for the 1000 realizations of each eBOSS DR16 sample. Each mock realization is handled as a set of independent data, and the $\chi^2$ minimization is performed on the residual $\epsilon(\beta)$ as defined by Eq.~\ref{eqn:residuals}. The covariance for the mock is computed with the $N_s - 1 = 999$ remaining mocks. The measurement of the correlation function is performed over the range $r/r_v=\left[ 0;3.6\right]$ with a number of 22 bins, 18 bins and 22 bins for LRG+CMASS, ELG and QSO samples respectively (see Section~\ref{sec:ana_syst_bin} for a detailed description of the optimization). The fitting procedure as described in Section~\ref{sec:ana_fit} is illustrated in Fig.~\ref{fig:Bestfit_mock} for one \EZ{} catalogue of each eBOSS sample. 

\begin{figure}
\begin{subfigure}{\columnwidth}
\caption{\EZ{} LRG+CMASS}
\includegraphics[width=\columnwidth]{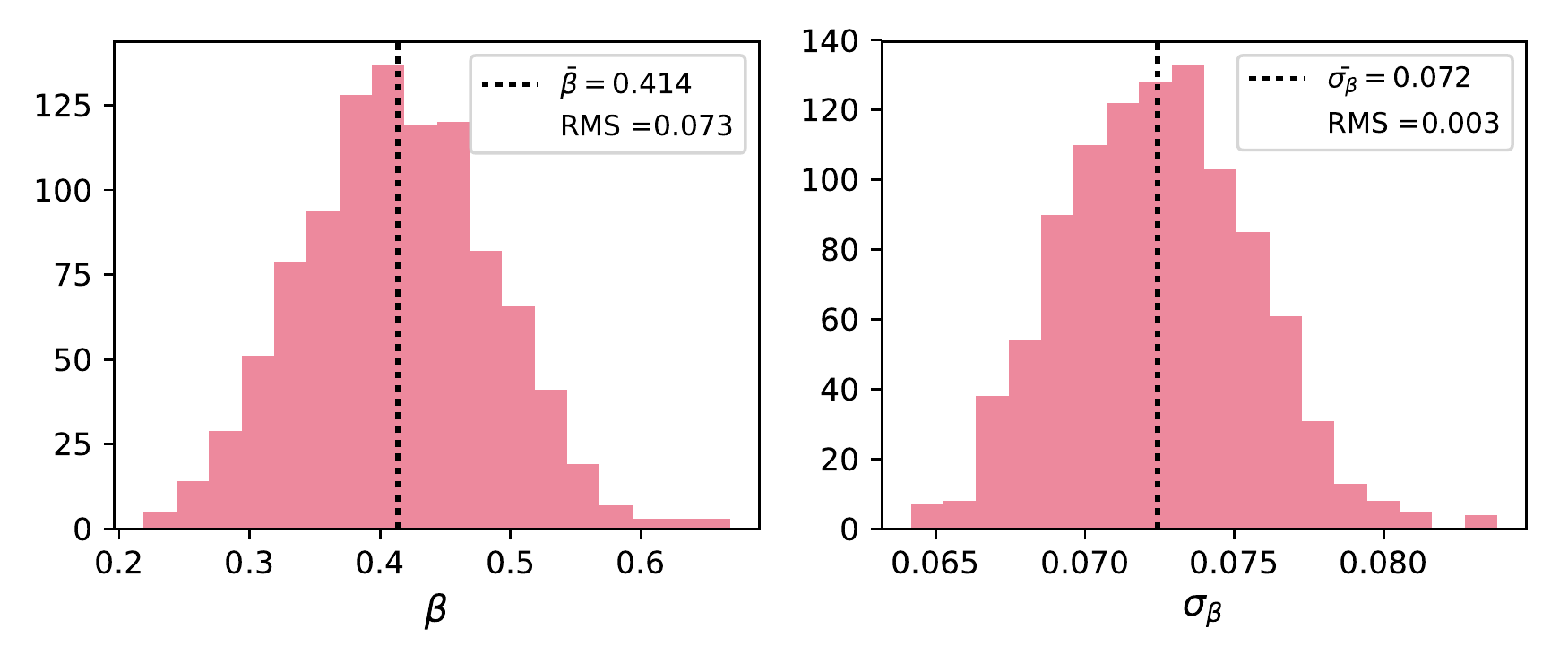}
\label{subfig:LRG+CMASS_be_mocks}
\end{subfigure}
\begin{subfigure}{\columnwidth}
\caption{\EZ{} ELG}
\includegraphics[width=\columnwidth]{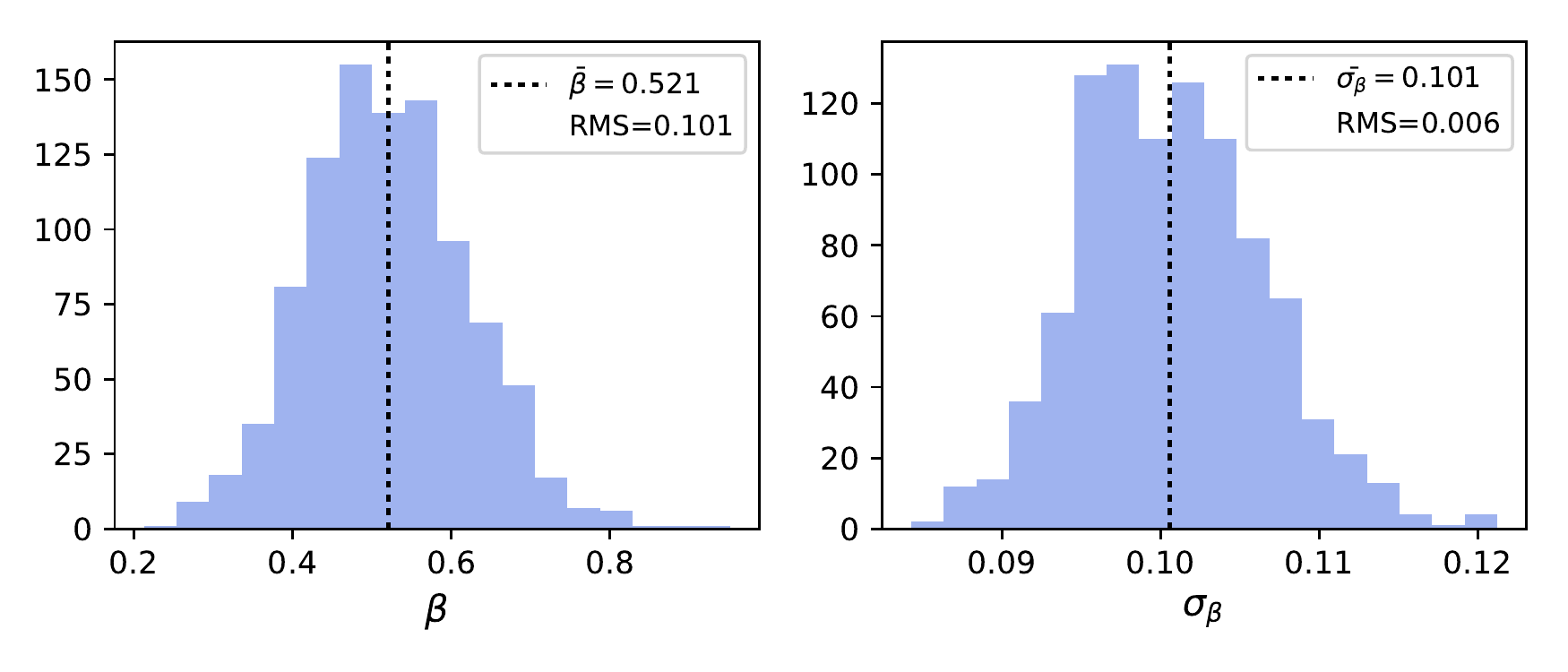}
\label{subfig:ELG_be_mocks}
\end{subfigure}
\begin{subfigure}{\columnwidth}
\caption{\EZ{} QSO}
\includegraphics[width=\columnwidth]{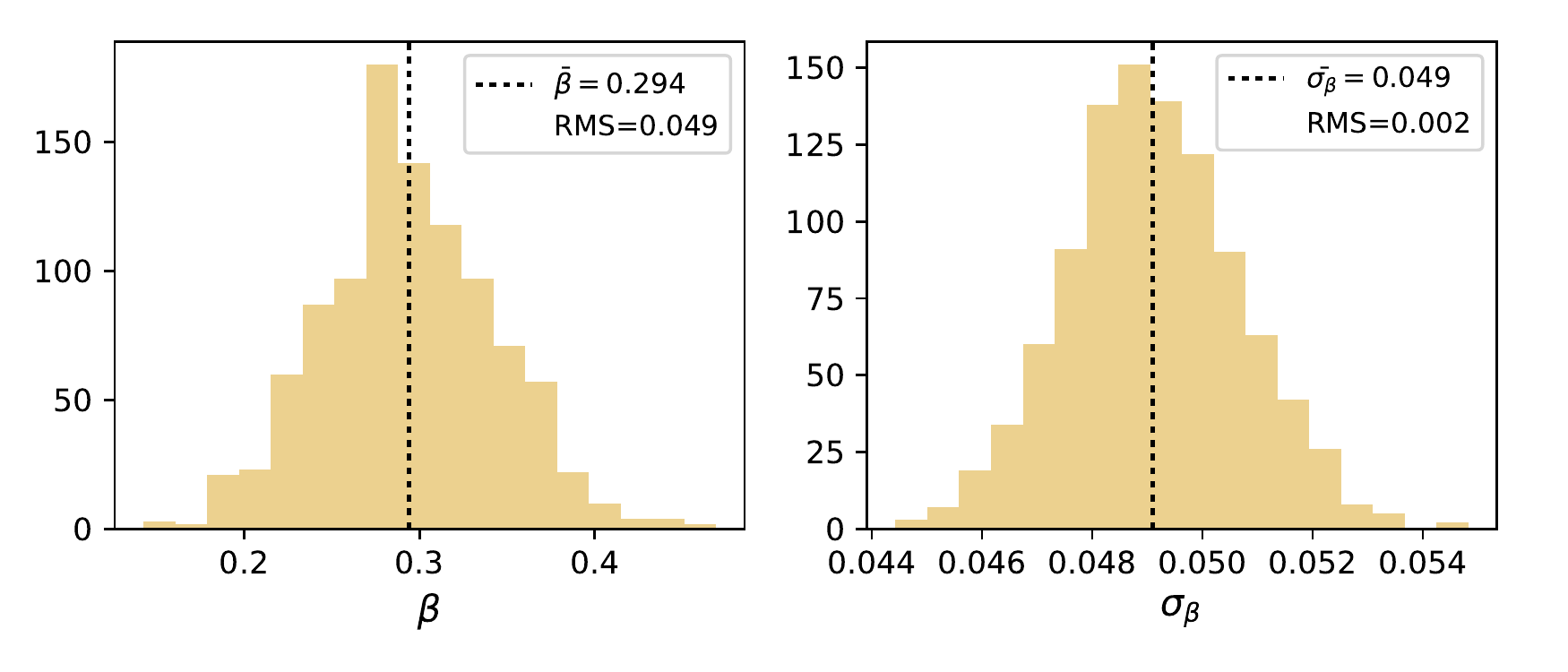}
\label{subfig:QSO_be_mock} 
\end{subfigure}
\caption{Best-fitting parameters for the 1000 realizations (950 for the ELG sample) of the \EZ{} catalogues. Left panels display the distribution of the distortion parameter $\beta$ and right panels display the distribution of the errors of $\beta$. The LRG+CMASS, ELG and QSO \EZ{} samples are displayed in the top (a), middle (b) and bottom (c) panels, respectively.} 
\label{fig:beta_mock}
\end{figure}

In Figure~\ref{fig:beta_mock} we display the recovered $\beta$ values from the 1000 \EZ{} realizations as well as the associated error. The RMS of the $\beta$ distribution from the 1000 \EZ{} (950 mocks for the ELG sample) is similar to the mean value of the $\sigma_{\beta}$ distribution, showing that the full distribution for $\beta$ follows a Gaussian distribution. The mean values of $\beta$, $\sigma_{\beta}$ and $\chi^2$ are reported in Table~\ref{table:mocks_ref} for each eBOSS tracer. 

\begin{table}
    \caption{Statistics on the distortion parameter fit on the 1000 \EZ{} (950 for the ELG sample) realizations for each eBOSS tracer. The error $\langle\sigma_{\beta}\rangle$ is the mean value of the individual fitting errors. The $\chi^2$ is normalized to the number of degrees of freedom. The quoted $\beta$ value is used as the reference value for systematic tests performed in Section~\ref{sec:ana_syst}. }
    \centering
    \begin{tabular}{lccc}
    \hline
    \EZ  & $\langle \beta_\textsf{ref} \rangle$ & $\langle\sigma_{\beta}\rangle$ & $\langle\chi^2\rangle$ \\
    \hline\hline
    LRG+CMASS & 0.414 & 0.072 & 1.39  \\
    ELG       & 0.521 & 0.101 & 1.14 \\
    QSO       & 0.294 & 0.049 & 1.76   \\
    \hline
    \hline
    \end{tabular}
    \label{table:mocks_ref}
\end{table}

\subsection{Systematic tests}
\label{sec:ana_syst}
In this section we aim to run our fitting procedure on \EZ{} and N-Body mocks in order to check potential systematic errors. For each test, we estimate the bias on the value of the distortion parameter $\beta$ relative to the expected value $\beta_\textsf{ref}$. We set the systematic value to the maximum contribution between the bias and the $1\sigma$ error on the bias measurement. As systematics can differ between each eBOSS tracer, we use as notation: 
\begin{eqnarray}
\sigma_{\textsf{syst}} = (\sigma^{\textsf{LRG}}_{\textsf{syst}}, \sigma^{\textsf{ELG}}_{\textsf{syst}}, \sigma^{\textsf{QSO}}_{\textsf{syst}}) .
\end{eqnarray}

\subsubsection{Optimal number of bins}
\label{sec:ana_syst_bin}
We first study the optimal number of bins used for the measurement of the correlation function and the fitting procedure to extract the redshift distortion parameter $\beta$. It is worth noting that the optimal number of bins is not necessarily the same for the three eBOSS DR16 samples, as the number of galaxies and the sky coverage are not the same. The fitting range goes from $r/r_v=0$ to $3.6$. Increasing the number of bins helps to better shape the monopole, but at the cost of a reduced signal-to-noise ratio.

In order to determine the optimal number of bins for each sample, we conducted the full pipeline analysis using different binning schemes, as summarized in Table~\ref{table:syst_bin}. The final number of bins selected is a compromise between minimizing the relative error on $\beta$ and minimizing the $\chi^2$ of the fit. The selected number of bins is 22, 18 and 22 for LRG+CMASS, ELG and QSO samples, respectively. The impact of the choice of the binning size on the $\beta$ parameter is also given in Table~\ref{table:syst_bin}, where the error reported for $\langle \beta \rangle$ is the rms divided by $\sqrt{1000}$.
The deviation is about $4.8\%$ for the LRG sample, $2.3\%$ for the ELG sample and $1.4\%$ for the QSO sample. To be conservative, we quote the highest shift as the systematic uncertainty due to the binning scheme in each sample : 
\begin{eqnarray}
\label{eqn:syst_bin}
\sigma_{\textsf{syst,bin}} = (0.020, 0.012, 0.004) .
\end{eqnarray}

\begin{table}
    \caption{Performance of the number of bins $N_b$ used for the fitting procedure. We display the relative error on $\beta$, the reduced $\chi^2$ and the shift of distortion parameter with respect to the $\beta$ reference values quoted in Table~\ref{table:mocks_ref}. The reported values between $\langle\rangle$ are the means of best-fit parameters from each 1000 \EZ{} realizations of each eBOSS tracer. The error on means is the rms divided by $\sqrt{1000}$. The final number of bins is a compromise between minimizing the relative error on $\beta$ and minimizing the average $\chi^2$. The final number of bins is indicated in bold.}
    \centering
    \begin{tabular}{lcccc}
    \hline
    \EZ & $N_b$   &  ${\langle\sigma_\beta\rangle} / {\langle\beta\rangle}$ & $\langle\chi^{2}\rangle$ & $\langle \beta \rangle-\langle \beta_\textsf{ref} \rangle$\\
    \hline\hline
    LRG+CMASS & 16 & 0.177 & 1.57 & $0.020\pm0.004$\\
    LRG+CMASS & 18 & 0.178 & 1.51 & $0.018\pm0.004$\\
    LRG+CMASS & 20 & 0.178 & 1.48 & $0.016\pm0.004$\\
    \textbf{LRG+CMASS} & \textbf{22} &  \textbf{0.180} & \textbf{1.39} & \textbf{-} \\
    LRG+CMASS & 25 & 0.179 & 1.40 & $0.012\pm0.004$\\
    \hline
    ELG  &  14   & 0.214  & 1.57 & $0.011\pm0.005$ \\
    ELG  &  16   & 0.215 & 1.53  & $0.001\pm0.005$ \\
    \textbf{ELG} &  \textbf{18} & \textbf{0.214} & \textbf{1.48} & \textbf{-}  \\
    ELG  &  20   & 0.217 & 1.45 & $-0.007\pm0.005$  \\
    ELG  &  22   & 0.219 & 1.42 & $-0.012\pm0.005$ \\
    \hline
    QSO  &  16 & 0.169  & 2.04  & $0.004\pm0.002$  \\
    QSO  &  18 & 0.170  & 1.98  & $0.001\pm0.002$ \\
    QSO  &  20  & 0.169 & 1.89  & $-0.001\pm0.002$\\
    \textbf{QSO}  &  \textbf{22} & \textbf{0.169} & \textbf{1.76} &  \textbf{-} \\
    QSO  &  25 & 0.170  & 1.66  & $-0.002\pm0.002$ \\
    \hline
    \hline
    \end{tabular}
    \label{table:syst_bin}
\end{table}

\subsubsection{FKP weight}
Various weights are applied to galaxies in order to correct for observational systematics of the survey. In contrast, the FKP weight is introduced to compensate for the non-uniform radial distribution of the galaxies with the aim of minimizing the variance at the BAO scale. In the case of voids, we are not concerned with the BAO constraint, and it seems legitimate to ask whether this weight should be used in our analysis, in particular in the calculation of the cross-correlation function. We have therefore studied the impact of using the FKP weight or not when recovering the distortion parameter $\beta$. The difference of the mean $\beta$ values calculated from the 1000 \EZ{} realizations with and without the $w_{\textsc{FKP}}$ are given in Table~\ref{table:syst_FKP} under the label 'no FKP weight' for each tracer. The resulting systematic uncertainty from FKP correction is:
\begin{eqnarray}
\label{eqn:syst_fkp}
\sigma_{\textsf{syst,FKP}} = (0.006, 0.012, 0.002) ,
\end{eqnarray}
giving a relative uncertainty about $1.4\%$, $2.3\%$ and $0.7\%$ for the LRG, ELG and QSO sample, respectively.

\begin{table}
    \caption{Performance of the FKP weight and correlation function estimator in the $\beta$ parameter. We report the shift of distortion parameter with respect to the $\beta$ reference value quoted in Table~\ref{table:mocks_ref}. The difference is computed between the means of best-fit parameters from each 1000 \EZ{} realizations of each eBOSS tracer. The error on means is rms divided by $\sqrt{1000}$.}
    \centering
    \begin{tabular}{llc}
    \hline
    \EZ  & syst & $\langle \beta \rangle - \langle \beta_\textsf{ref} \rangle$  \\
    \hline\hline
    LRG+CMASS  & no FKP weight & $0.006\pm0.005$  \\
    LRG+CMASS  & LS estimator  & $-0.009\pm0.004$  \\
    \hline
    ELG & no FKP weight & $0.012\pm0.005$ \\
    ELG & LS estimator  & $0.017\pm0.005$ \\
    \hline
    QSO & no FKP weight  & $0.001\pm0.002$  \\
    QSO & LS estimator  & $0.003\pm0.002$  \\
    \hline
    \hline
    \end{tabular}
    \label{table:syst_FKP}
\end{table}

\subsubsection{Estimator}
The reasons why we use the DP-estimator (Eq.~\ref{eq:PD})  and not the LS-estimator (Eq.~\ref{eq:LS}) for the calculation of the void-galaxy cross-correlation function are given in Section~\ref{sec:ana_cor}. 
Nevertheless, these estimators have different properties of bias and variance~\citep{Vargas2013}. In this section we investigate a simplified LS-estimator that does not use the term $R_v$, as defined in~\citet[]{Hamaus2017}:  
\begin{eqnarray}
\xi^{LS}(r,\mu) \approx D_vD_g-D_vR_g .
\end{eqnarray}
The comparison on the $\beta$ mean value calculated from the 1000 \EZ{} realizations between the LS-estimator and the DP-estimator is shown in Table~\ref{table:syst_FKP} under the label 'LS estimator'. The effect is about $2.2\%$ for LRG+CMASS, $3.3\%$ for ELG and $1\%$ for QSO. The resulting systematic error associated with the choice of estimator is:
\begin{eqnarray}
\label{eqn:syst_estimator}
\sigma_{\textsf{syst,LS}} = (0.009, 0.017, 0.003) .
\end{eqnarray}

\subsubsection{RSD linear modelling}
\label{sec:sys_RSD}
In order to validate the RSD modelling, we performed the full analysis using N-body simulations that are supposed to predict as accurately as possible the expected RSD in the signal. 

We use the $N_{s}=84$ \NSeries{} mocks, the $N_{s}=30$
\OR{} ELG mocks and the $N_{s}=100$ \OR{} QSO mocks for the LRG, ELG and QSO samples, respectively, as described in Section~\ref{sec:ana_mock}. For each realization, we compute the cross-correlation function and its multipoles and fit the distortion parameter $\beta$ using the covariance matrix from the $N_s$ realizations. The best-fitting values for $\beta$ and $\sigma_\beta$ are summarized in Table~\ref{table:nseries_ref} for each eBOSS tracer. 

In order to validate our RSD model, we compare the recovered value of the distortion parameter $\beta^\textsf{NB}$ with the fiducial $\beta^\textsf{fid}$ value of each set of simulations. The fiducial $\beta^\textsf{fid}$ values are defined as the ratio $f/b$, where $f$ is derived from the fiducial cosmology as given in Section~\ref{sec:ana_mock} and where the galaxy bias $b$ is provided by the DR16 companion papers for the LRGs~\citep{Bautista20,Gilmarin20}, ELGs
~\citep{Tamone20,DeMattia20} and QSOs~\citep{Hou20,Neveux20}. Our results show that deviations are larger than $1\sigma$ error as quoted in the last column of the Table~\ref{table:nseries_ref}, where the $1\sigma$ error is the rms divided by $\sqrt{N_s}$. The relative difference is about $9\%$, $8\%$ and almost $40\%$ compared to the fiducial values. The discrepancy for the QSO sample is surprisingly large, and not well understood at this stage. However we adopt a conservative approach, and consider this discrepancy to be a systematic error.

\begin{table}
    \caption{Performance of the RSD modelling. We display results on the distortion parameter fit for the \NSeries{} LRG ($N_s=84$ realizations), the \OR{} ELG ($N_s=30$ realizations) and the \OR{} QSO ($N_s=100$ realizations) simulations. The error $\langle\sigma_{\beta}\rangle$ is the mean value of the individual fitting errors. Fiducial values $\beta^\textsf{fid}$ for these N-body simulations are defined as the ratio $f/b$, where $f$ is derived from the fiducial cosmology as given in Section~\ref{sec:ana_mock} and the galaxy bias $b$ is given by the DR16 companion papers~\citep{Bautista20,Gilmarin20,Tamone20,DeMattia20,Hou20,Neveux20}. The last column gives an estimate of the measured bias due to the RSD modelling, where the error is the rms divided by the squared root of the number of mocks $\left( \frac{\langle\sigma_{\beta}\rangle}{\sqrt{N_s}} \right)$.}
    \centering
    \begin{tabular}{lcccc}
    \hline
       & $\langle \beta^\textsf{NB}_\textsf{ref} \rangle \pm \langle\sigma_{\beta}\rangle$ & $\beta^\textsf{fid}$ & $\langle \beta \rangle -  \beta^\textsf{fid}$\\
    \hline\hline
    \NSeries{} LRG & 0.447 $\pm$ 0.063 &  0.41 & $0.037\pm0.007$ \\
    \OR{} ELG   & 0.629 $\pm$ 0.027 &  0.686 & $0.057\pm0.005$\\
    \OR{} QSO   & 0.241 $\pm$ 0.037&  0.401  & $0.160\pm0.004$\\
    \hline
    \hline
    \end{tabular}
    \label{table:nseries_ref}
\end{table}

\subsubsection{Fiducial cosmology}
\label{sec:syst_fid}
The void finding algorithm needs to convert galaxy redshifts into distance in order to perform tesselation and define voids. It therefore requires a fiducial cosmology parametrized by the value $\Omega_m^\textsf{fid}$ as input. In this section, we study  the systematic error introduced by this choice. 

For this study we used the \NSeries{} mocks whose the true cosmology is $\Omega_m^\textsf{true}=0.286$. We conducted our study using two different fiducial cosmologies, the first with $\Omega_m^\textsf{fid}=\Omega_m^\textsf{true}=0.286$, and the second with $\Omega_m^\textsf{fid}=0.31$. These fiducial cosmologies are used both in the void finder and in the calculation of the correlation function. 
Table~\ref{table:nseries_syst} displays results on the recovered $\beta$ parameter using both cosmologies, under the label 'barycentre' which is our baseline for the void center definition (see discussions about void center definition in Section~\ref{sec:syst_centre}). The reference $\beta^\textsf{NB}_\textsf{ref}$ value is taken from Table~\ref{table:nseries_ref}. We find that the bias on the recovered parameter is negligible, of the order of $0.7\%$, and is dominated by its error, which is quite large due to the low number of mock used. 
We take the $1\sigma$ error on the deviation measurement to be the systematic error associated with the choice of fiducial cosmology 
\begin{eqnarray}
\label{eqn:syst_fid}
\sigma_{\textsf{syst,fid}} = 0.010,
\end{eqnarray} 
corresponding to a $2.2\%$ effect.

\begin{table}
    \caption{Performance of the fiducial cosmology and definition of the void centre on the \NSeries{} mocks. We report the shift of distortion parameter with respect to the $\beta$ reference value quoted in Table~\ref{table:nseries_ref} and which refers to the first row (our baseline). The difference is computed between the means of best-fit parameters over the 84 \NSeries{} realizations. The error on means is rms divided by $\sqrt{84}$.}
    \centering
    \begin{tabular}{lccc}
    \hline
    \NSeries  & $\Omega_m^\textsf{fid}$ & $\langle \beta \rangle - \langle \beta^\textsf{NB}_\textsf{ref} \rangle$  \\
    \hline\hline
    barycentre      & 0.286 & - \\
    barycentre      & 0.31 & $0.003\pm0.010$ \\
    \hline
    circumcentre    & 0.286 & $0.018\pm0.010$  \\
    circumcentre    & 0.31 & $0.079\pm0.010$  \\
    \hline
    \hline
    \end{tabular}
    \label{table:nseries_syst}
\end{table}

\subsubsection{Void centre definition}
\label{sec:syst_centre}
When calculating the void-galaxy cross-correlation function defined by Eq.~\ref{eq:PD}, the separation distance is measured from the centre of the considered void. Now, in the \textsc{Revolver} void finder, we can use two different definitions of the void centre: the \textit{barycentre}, defined as the arithmetic mean of the coordinates of galaxies weighted by their Voronoi volume (see Section~\ref{sec:void_finding}), and the \textit{circumcentre}, that is computed from the four lowest density Voronoi cells. We justify here our choice of the void center definition. 

Table~\ref{table:nseries_syst} displays results on the distortion parameter $\beta$ using the barycentre (our default) or the circumcentre definition. The values of the recovered $\beta$ parameter are given for both fiducial cosmologies studied in Section~\ref{sec:syst_fid}. If we only consider the bias induced by the choice of the void centre definition in the case of $\Omega_m^\textsf{fid}=\Omega_m^\textsf{true}$, then the effect is of the order of $4\%$. However, we report a significant deviation in the case of $\Omega_m^\textsf{fid}\ne\Omega_m^\textsf{true}$, meaning that the definition of barycenter is more robust to fiducial cosmology than that of the circumcentre of the voids. This gives us confidence in the choice of the barycentre for our baseline, and as such, we do not attribute any systematic error to the choice of the voidcentre definition.

\subsubsection{Buffer density ratio} 
As mentioned in Section ~\ref{sec:void_stat}, the \textsc{Revolver} algorithm was run 1000 times on the data catalogues, in order to minimize the inherent dispersion due to the random positioning of buffer particles that can impact the positions and properties of voids. Here we evaluate the systematic error related to this procedure. 

For this purpose we apply 1000 times the void finder on the same \EZ{} catalogue. This catalogue is arbitrarily chosen among the 1000 available. The associated systematic error is not the bias on the measurement, but the error on the average value of the $\beta$ recovered from fitting each individual mock. The rms of the $\beta$ distribution rounds up to $0.015$. With 1000 realizations of the void catalogue, the error becomes negligible, less than $\delta\beta=5.10^{-4}$. We also check that we recover these values when fitting the data (see Section
~\ref{sec:results}). Therefore we consider this effect to have a negligible contribution to the total systematic error budget.

\subsubsection{Systematic error budget}
In this section we summarize the error budget. As the errors are dependent on the mocks used, we summarize in Table~\ref{table:total_syst} the list of \textit{relative} systematic contributions, which will allow us to rescale them to the value of the $\beta$ measured in the data. Contributions can be classified into three categories, the dominant effect coming from the validation of the RSD modelling. 
Finally, the total relative systematic error is the quadratic sum of each contribution is: 
\begin{eqnarray}
\label{eqn:syst_tot}
\sigma_{\textsf{syst,tot}} = (10.8\%, 9.8\%, 40\%).
\end{eqnarray}

\begin{table}
    \caption{Summary of systematic relative errors on the $\beta$ parameter obtained from tests with mock catalogues for each of the eBOSS tracer. The total systematic error is the quadratic sum of each contribution.}
    \centering
    \begin{tabular}{llccc}
    \hline
    Type & systematics in $(\sigma_\beta/\beta)\;(\%)$ & ${\textsf{LRG}}$ & ${\textsf{ELG}}$ & ${\textsf{QSO}}$  \\
    \hline\hline
    Correlation & Binning  &  4.8 &  2.3 &  1.4  \\
    function & FKP weight  &  1.4 &  2.3 &  0.7  \\
             & Estimator   &  2.2 &  3.3 &  1.0 \\
    \hline
    Void    \\
    finder & Fiducial cosmology  &  2.2 &  2.2 &  2.2 \\
    \hline
    Model & RSD modelling & 9.0 & 8.3 & 39.9\\

    \hline
    Total ($\%$) &  & 10.8 & 9.76 & 40.0  \\
    \hline \hline
    \end{tabular}
    \label{table:total_syst}
\end{table}

\section{Results}
\label{sec:results}
In this section we apply the fitting procedure optimized with \EZ{} on the final release of eBOSS, the DR16 dataset. We present our measurements in terms of the distortion parameter $\beta$ (Section~\ref{sec:res_beta}). Then, in order to compare our results with the literature, we explain how we convert our $\beta$ measurements in terms of constraints on the growth rate of structure (Section~\ref{sec:res_fsig}). 

\subsection{Measurements of the distortion parameter $\beta$}
\label{sec:res_beta}
\begin{figure}
\begin{subfigure}{\columnwidth}
\includegraphics[width=\columnwidth]{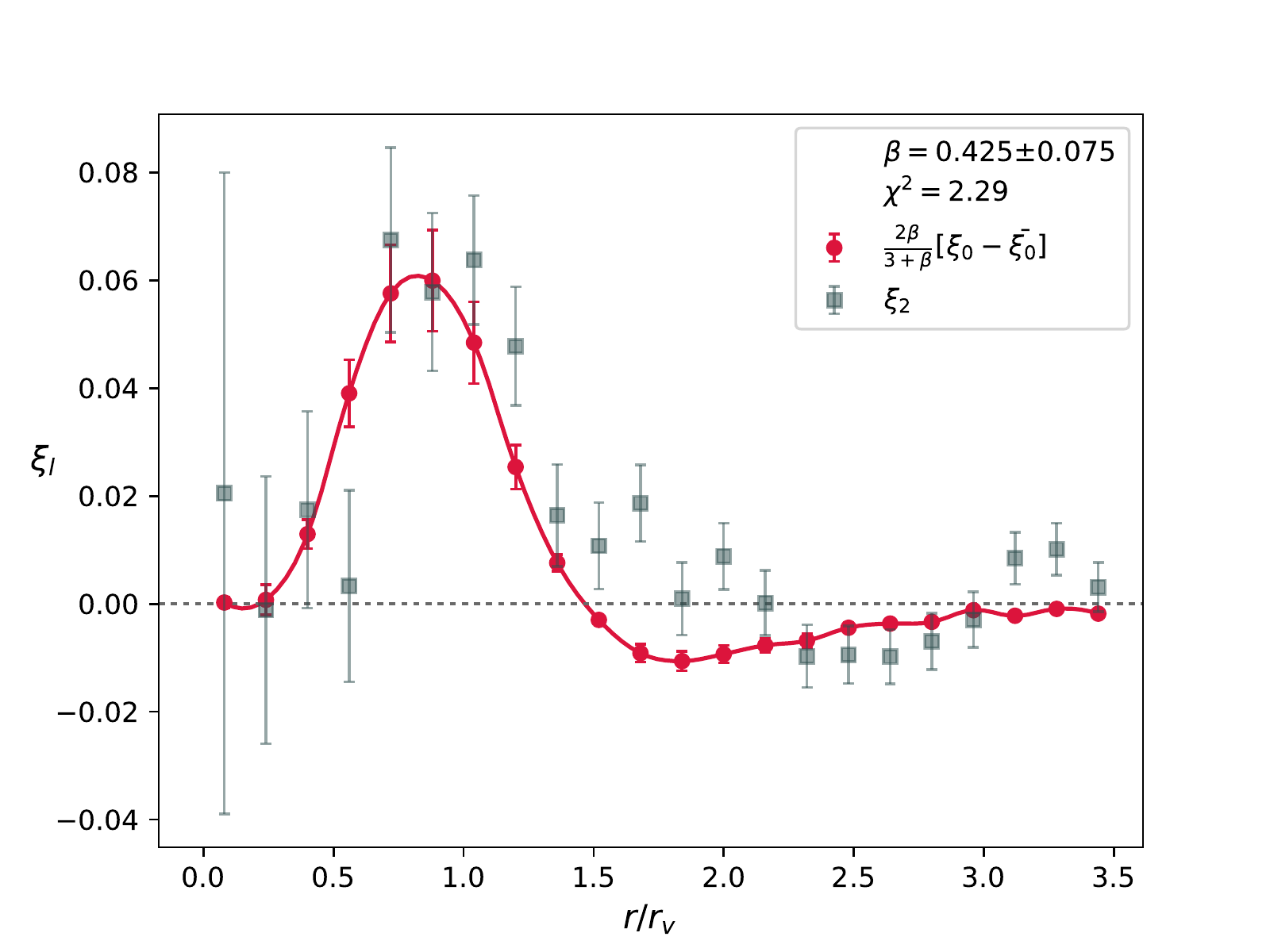}
\caption{DR16 LRG+CMASS}
\label{subfig:LRG_Bestfit} 
\end{subfigure}
\begin{subfigure}{\columnwidth}
\includegraphics[width=\columnwidth]{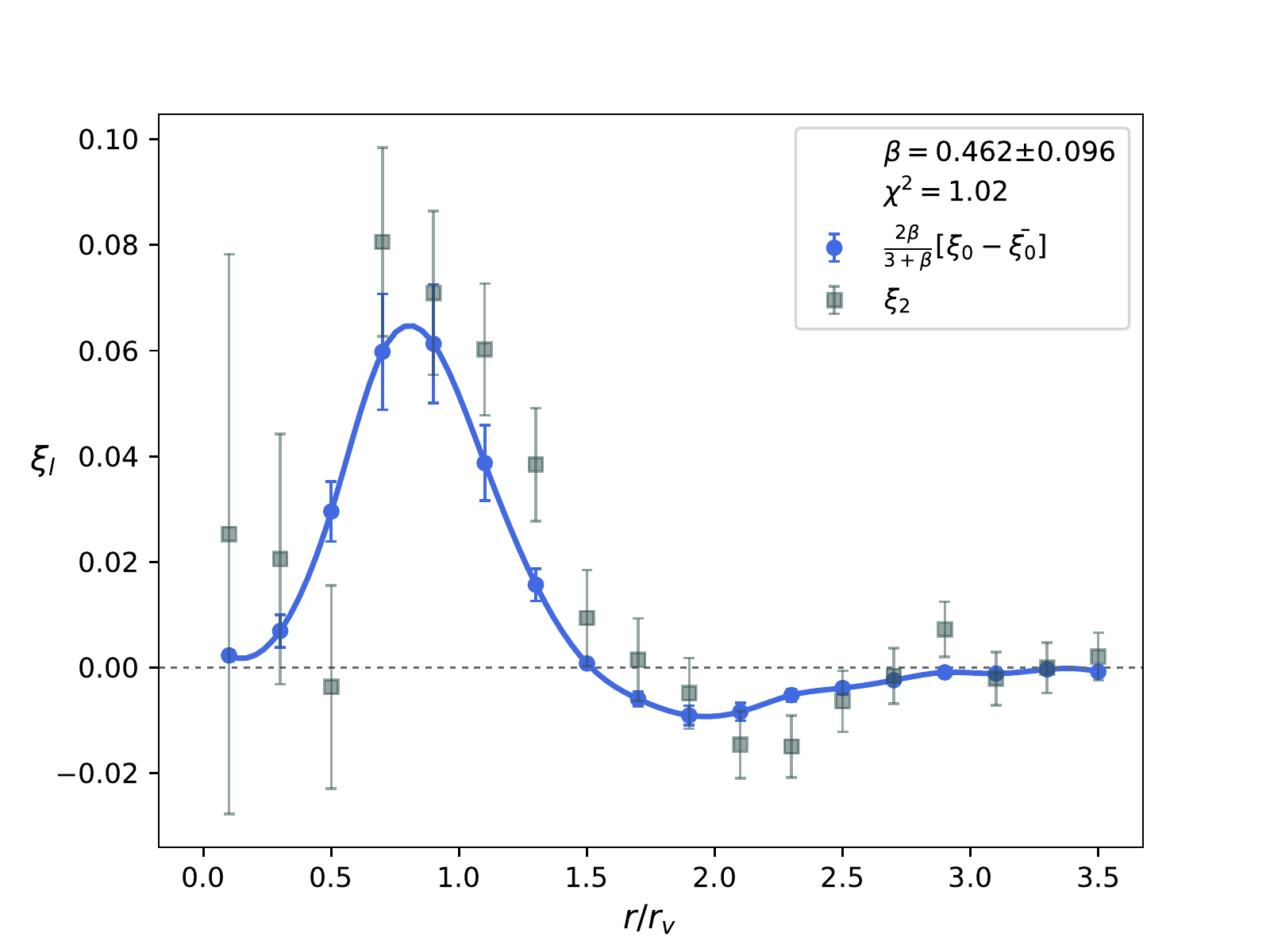}
\label{subfig:ELG_Bestfit}
\caption{DR16 ELG}
\end{subfigure}
\begin{subfigure}{\columnwidth}
\includegraphics[width=\columnwidth]{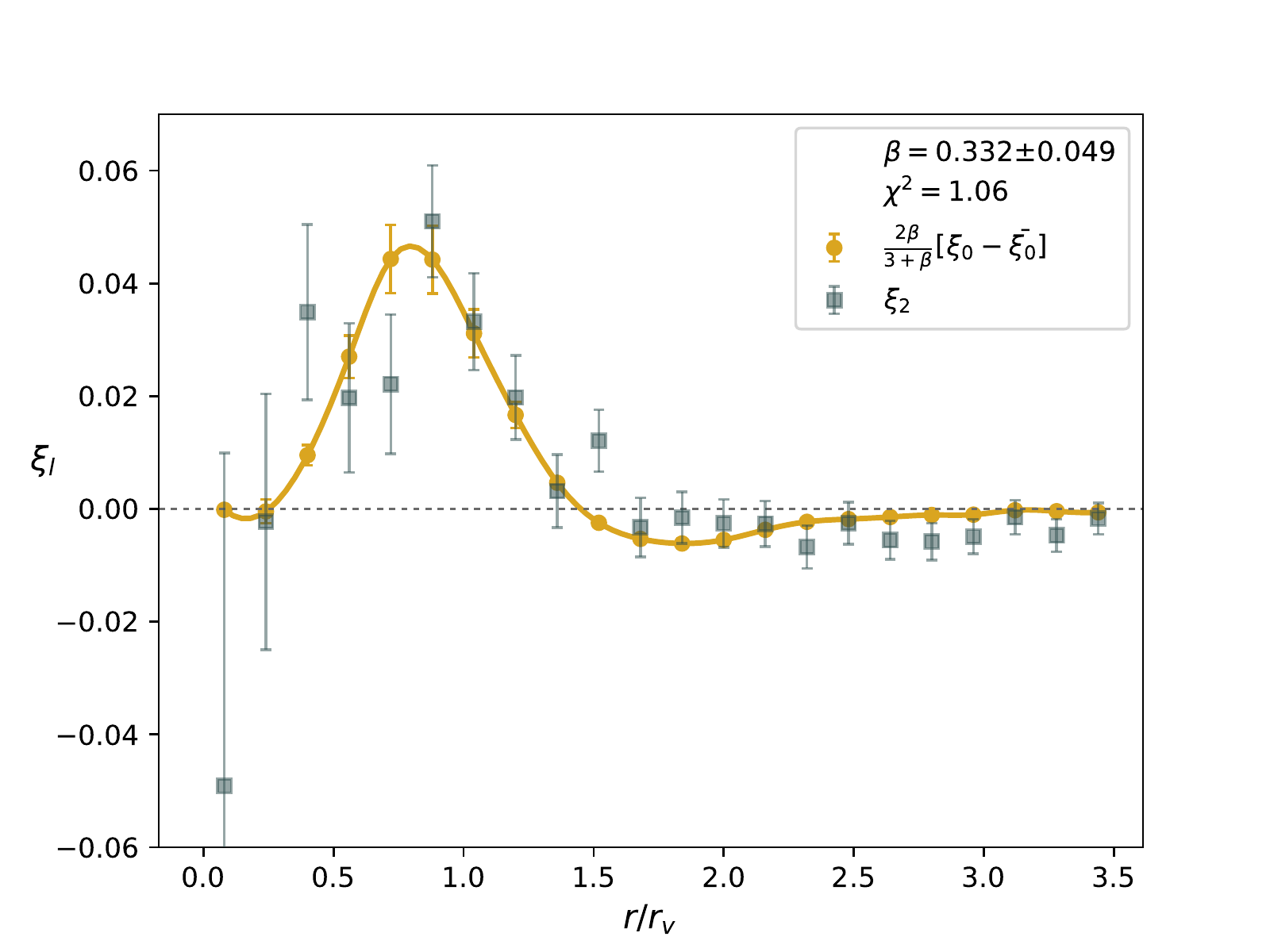}
\caption{DR16 QSO }
\label{subfig:QSO_Bestfit} 
\end{subfigure}
\caption{Quadrupole ($\xi_2$) and the best-fit of the  $ 2 \beta / (3 + \beta) (\xi_0-\bar{\xi_0})$ from one DR16 data catalogue of the LRG+CMASS, ELG and QSO sample displayed in the top (a), middle (b) and bottom (c) panels, respectively. Error bars are the diagonal of the covariance matrix from the 1000 \EZ{} realizations.}
\label{fig:Bestfit_data}
\end{figure}

\begin{figure}
\begin{subfigure}{\columnwidth}
\caption{DR16 LRG+CMASS}
\includegraphics[width=\columnwidth]{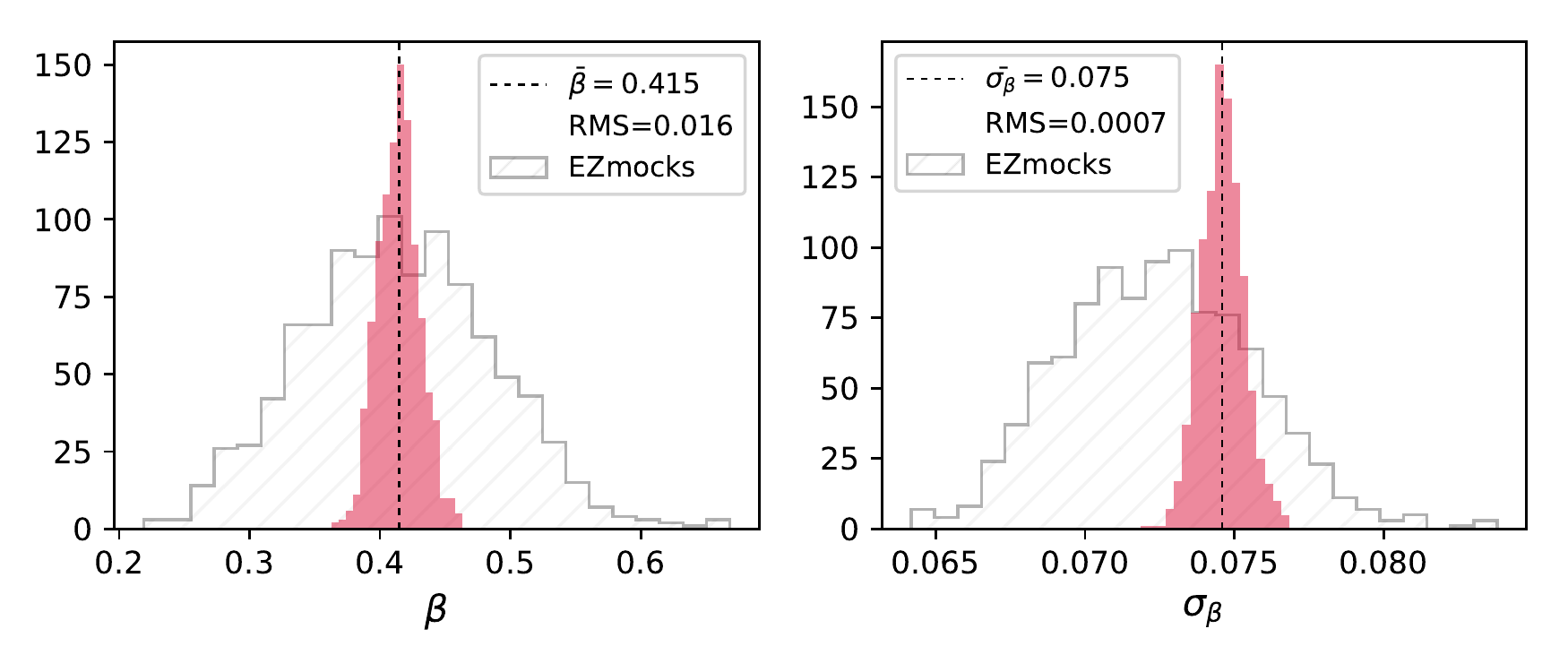}
\label{subfig:LRG_be_data}
\end{subfigure}
\begin{subfigure}{\columnwidth}
\caption{DR16 ELG}
\includegraphics[width=\columnwidth]{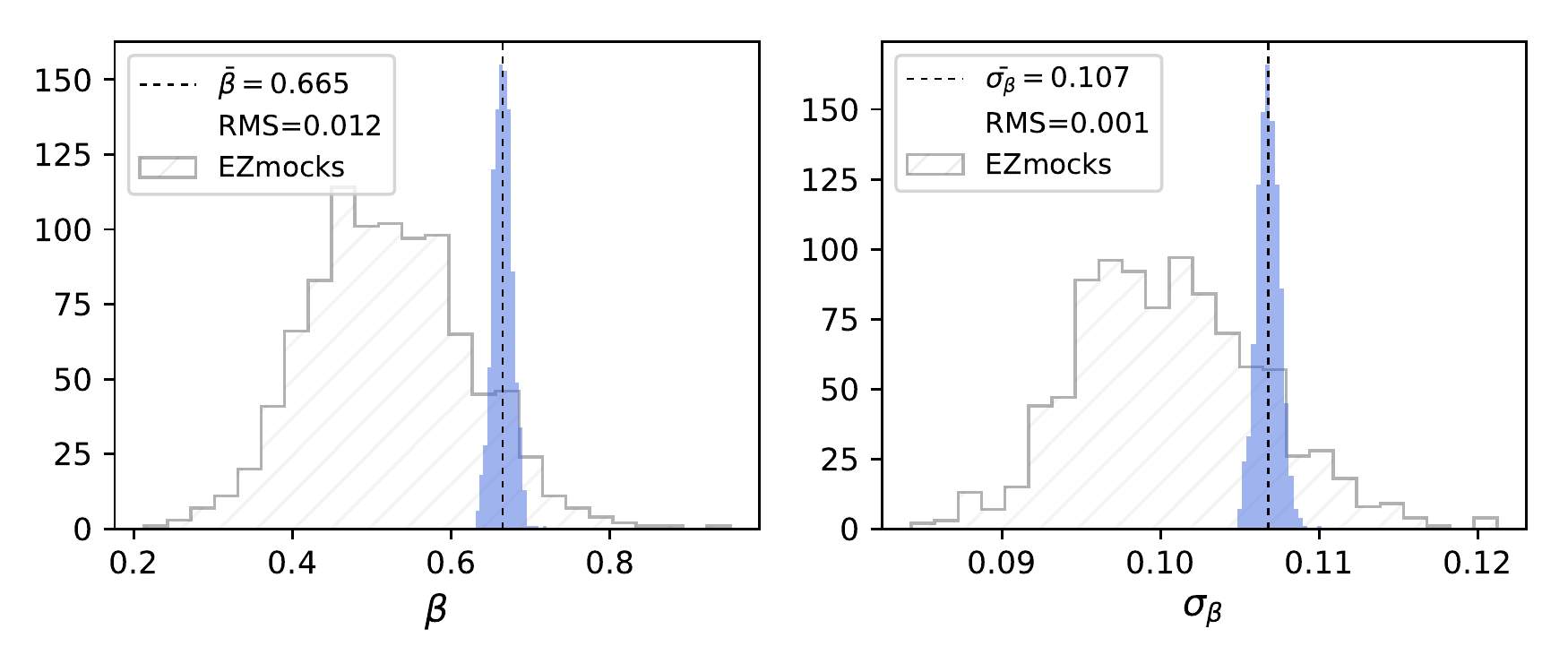}
\label{subfig:ELG_be_data}
\end{subfigure}
\begin{subfigure}{\columnwidth}
\caption{DR16 QSO}
\includegraphics[width=\columnwidth]{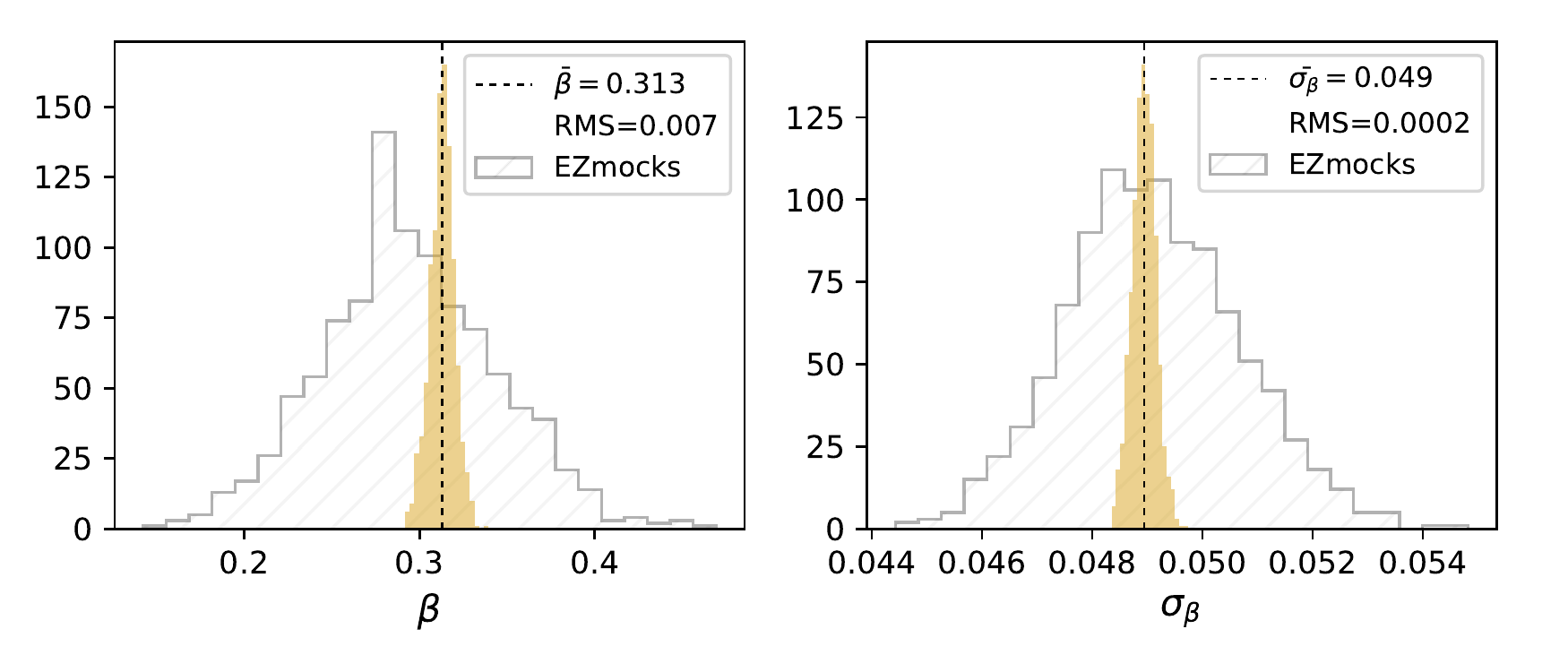}
\label{subfig:QSO_be_data} 
\end{subfigure}
\caption{Best-fitting parameters for the 1000 DR16 data catalogues respective to each tracer. Left panels display the distribution of the distortion parameter $\beta$ and right panels display the distribution of the errors of $\beta$. The LRG, ELG and QSO data samples are displayed in full color in the top (a), middle (b) and bottom (c) panels, respectively. For comparison the distribution of $\beta$ and $\sigma_{\beta}$ from \EZ{} is drawn in dashed regions.} 
\label{fig:beta_data}
\end{figure}

Figure~\ref{fig:Bestfit_data} displays the multipoles of the cross-correlation function and best-fit of the distortion parameter $\beta$ for one void catalogue of each eBOSS DR16 data sample. The covariance is computed from the 1000 \EZ{} realizations. The recovered $\beta$ values from the 1000 void catalogues are presented in Figure~\ref{fig:beta_data} for each eBOSS tracer. We note that the dispersion of $\beta$ is very small in comparison to that obtained from the 1000 \EZ{}, since the latter are dominated by the dispersion due to cosmic variance. The error on the mean value of $\beta$ is indeed the mean of the individual fit errors.

Final results on the distortion parameter $\beta$ are presented in Table~\ref{table:data_ref} for the three eBOSS DR16 datasets. The displayed statistical error is the mean value of the error $\sigma_\beta$ and the displayed systematic error is the relative error from Table~\ref{table:total_syst} renormalized to the measured $\langle \beta \rangle$ value. The total error $\sigma_\textsf{tot}$ is the quadratic sum of statistical and systematic errors.

\begin{table}
    \caption{Final results on the distortion parameter from the eBOSS DR16 void datasets. Mean values are recovered from the 1000 void catalogues generated from each eBOSS tracer. The quoted statistical error is the mean value of the error in the distortion parameter fit and the quoted systematic error is the total error given in Table~\ref{table:total_syst}. The total error is a quadratic sum of statistical and systematic errors. }
    \centering
    \begin{tabular}{lcccc}
    \hline
    Data samples & $\langle \beta \rangle$ & $\sigma_\textsf{stat}$ & $\sigma_\textsf{syst}$ & $\sigma_\textsf{tot}$ \\
    \hline\hline
    LRG       & 0.415 & 0.075 &  0.045 & 0.087 \\
    ELG       & 0.665 & 0.107 &  0.065 & 0.125\\
    QSO       & 0.313 & 0.049 &  0.125 & 0.134\\
    \hline
    \hline
    \end{tabular}
    \label{table:data_ref}
\end{table}

\subsection{Estimate of the growth rate $f\sigma_8$}
\label{sec:res_fsig}
The final growth rate measurement is obtained by combining $\beta$ and the linear bias $b_1$ according to: $f(z)=\beta b_1(z)$. However, as the galaxy bias is measured with a fixed normalization of $\sigma_8$, where $\sigma_8$ is the rms mass fluctuation in spheres with radius 8\hmpc, the measured value of $b_1$ is degenerate with $\sigma_8$. One way to be independent of this problem is to present our results in terms of $f(z)\sigma_8(z)$ as proposed by \citet[]{Song2009}, following:
\begin{eqnarray}
\label{eqn:fsigma8}
f\sigma_8 = \beta b_1\sigma_8 .
\end{eqnarray}

The measurement of $b(z)\sigma_8(z)$ is provided from galaxy clustering measurement through the estimate of the galaxy auto-correlation function. 
Because the underlying galaxy data are the same, we take here the measured values from the DR16 dataset with the clustering analyses conducted by companion papers: for the LRG+CMASS sample, BAO and RSD analyses were performed in configuration space~\citep{Bautista20} and Fourier space~\citep{Gilmarin20}; for the ELG sample, the galaxy clustering analyses in configuration space and in Fourier space are discussed in~\citet{Tamone20} and \citet{DeMattia20}, respectively; for the QSO sample, the quasar clustering is measured from the auto-correlation function~\citep{Hou20} and the power spectrum~\citep{Neveux20}. The corresponding $b_1\sigma_8$ values are presented in Table~\ref{table:result_fsig8}. We also report $\beta$ values from our analysis using voids, with the total error as quoted in Table~\ref{table:data_ref}. The resulting constraint on $f\sigma_8$ is given in the last column of Table~\ref{table:result_fsig8}, where the error includes the galaxy bias error contribution. We checked that $\beta$ and $b_1\sigma_8$ are slightly (anti-)correlated, meaning that we overestimated our error. 

\begin{figure*}
\begin{subfigure}{2\columnwidth}
    \includegraphics[width=\columnwidth]{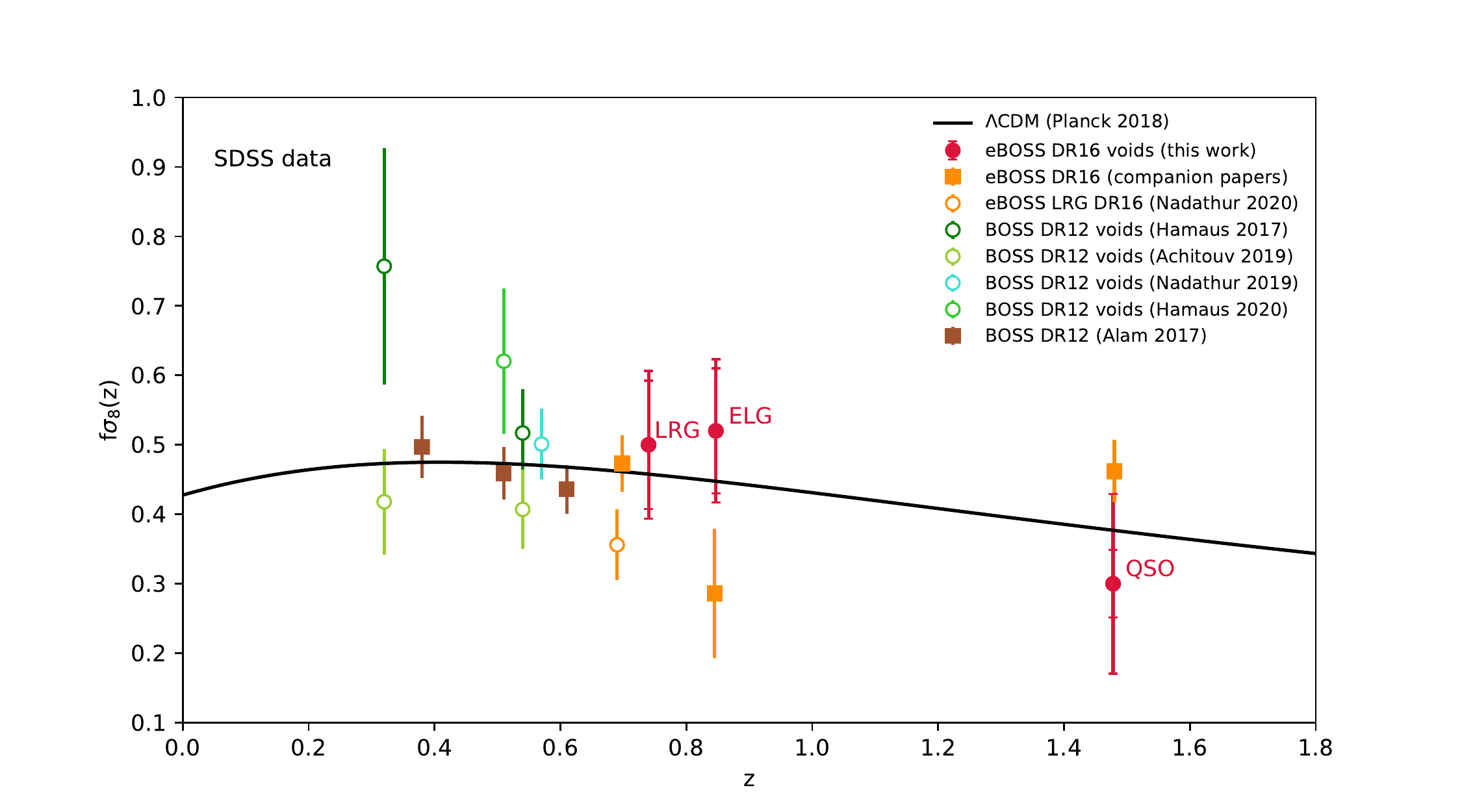}
    \label{fig:fsig8_sdss} 
\end{subfigure}
\begin{subfigure}{2\columnwidth}
    \includegraphics[width=\columnwidth]{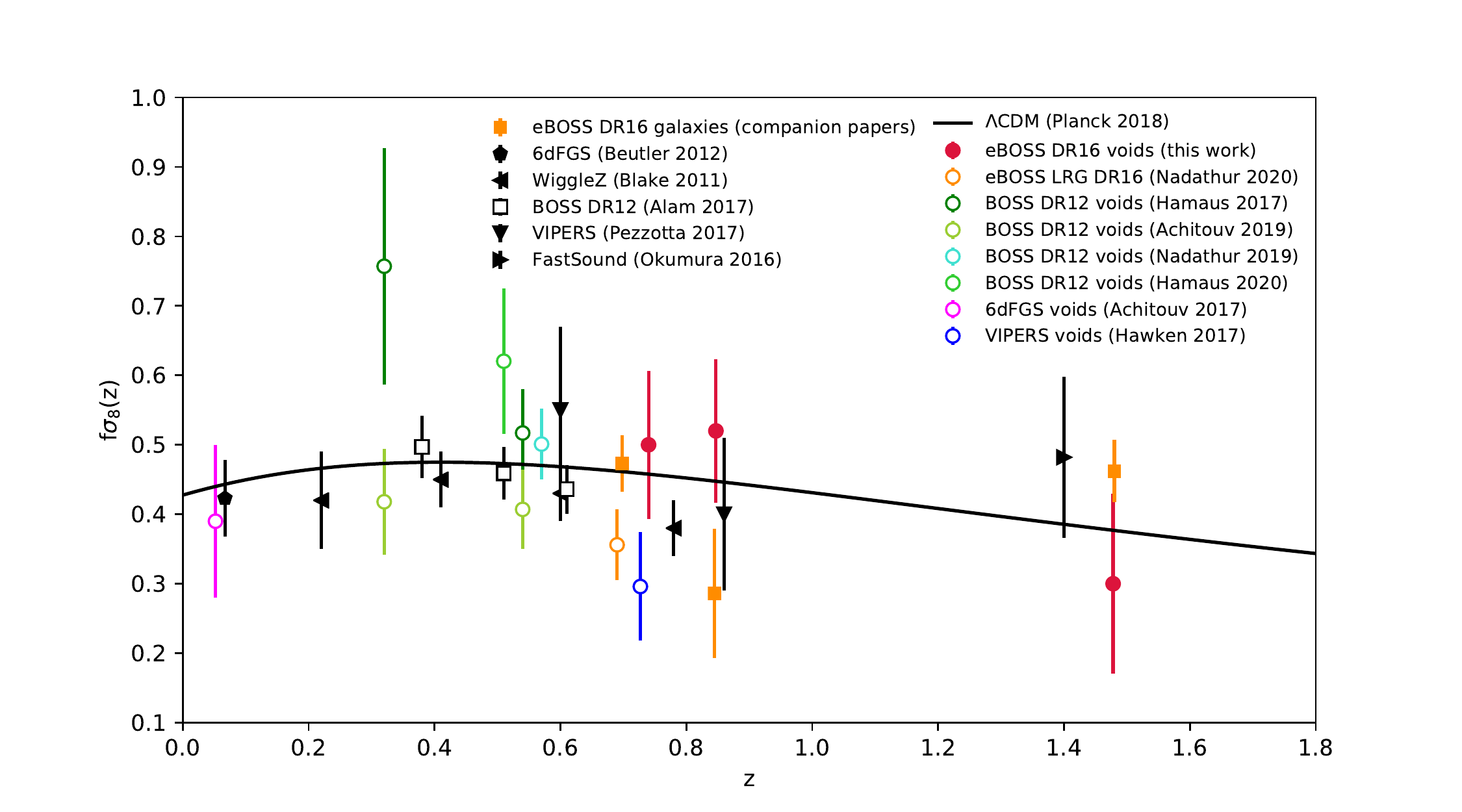}
    \label{fig:fsig8_world} 
\end{subfigure}
\caption{Comparison of $f\!\sigma_8(z)$ results to other measurements. 
Top panel shows the comparison with other estimates from SDSS data. The $f\!\sigma_8$ results from this work (red circles) are compared to constraints using voids (open circles) and conventional clustering techniques (filled squares) from eBOSS DR16 and BOSS DR12. For our measurements, we display the error contribution resulting from the RSD modelling uncertainty only by the outer error bars between caps. For DR16 datasets, we display the final consensus results (orange squares) from the LRG+CMASS sample~\citep[][]{Bautista20,Gilmarin20}, the ELG sample~\citep[][]{Tamone20,DeMattia20} and the QSO sample~\citep[][]{Hou20,Neveux20} to be compared to LRG voids, ELG voids and QSO voids, respectively. The constraint from a complementary void analysis performed on the eBOSS DR16 LRG+CMASS \citep{Nadathur2020-eboss} is also displayed. For DR12 datasets, we report $f\!\sigma_8$ measurements from galaxy clustering in BOSS~\citep[brown squares: ][]{Alam2017} with results from voids~\citep[open green and turquoise circles: ][]{Hamaus2017,Achitouv2019,Nadathur2019,Hamaus2020}.  
Bottom panel shows the comparison of $f\!\sigma_8$ results from this work (red circles) with other measurements using voids, in 6dFGS~\citep[open magenta circle: ][]{Achitouv2017}, in VIPERS~\citep[open dark blue circle: ][]{Hawken2017} and in BOSS DR12~\citep[open green and turquoise circles: ][]{Hamaus2017,Achitouv2019,Nadathur2019,Hamaus2020}. We also compare with conventional clustering measurements in the 6dFGS~\citep{Beutler2012}, the WiggleZ~\citep{Blake2011}, the BOSS DR12~\citep{Alam2017}, the VIPERS~\citep{Pezzotta2017} and  the FastSound~\citep{Okumura2016} surveys. We report results from the eBOSS DR16 companion papers (orange squares, see references above).
We overplot predictions for flat $\Lambda$CDM cosmological model assuming $\Omega_m = 0.31$ and $\sigma_8=0.81$.}
\label{fig:fsig8} 
\end{figure*}

Next we compare our $f\sigma_8$ results to those from the literature. The top panel of Figure~\ref{fig:fsig8} shows the comparison with work done within the SDSS Collaboration. Results from our work (red circles) are compared to the final consensus $f\sigma_8$ results from eBOSS DR16 using conventional clustering techniques (orange squares) for the LRG+CMASS sample~\citep{Bautista20,Gilmarin20}, ELG sample~\citep{Tamone20,DeMattia20} and QSO sample~\citep{Hou20,Neveux20}
and using voids (orange open circle) in the LRG+CMASS sample~\citep{Nadathur2020-eboss}.
We can note a slight shift in the effective redshift of the LRG+CMASS samples: This offset was caused by the selection cuts applied in our void catalogue, which mostly removed voids close to $z=0.6$. The error contribution resulting from the RSD modelling uncertainty in our measurement is highlighted by the outer error bars between caps. The agreement between galaxy clustering and void clustering is good, at the level of 1\,$\sigma$ for the three LRG, ELG and QSO samples. 

We also display in Figure~\ref{fig:fsig8} the $f\sigma_8$ results 
at lower redshift from BOSS DR12. These results include direct measurements from conventional galaxy clustering~\citep{Alam2017}, as well as $f\sigma_8$ constraints using voids~\citep[]{Hamaus2017,Achitouv2019,Nadathur2019, Hamaus2020}. In \citet[]{Nadathur2019} and \citet{Nadathur2020-eboss}, the authors performed a joint fit for redshift space distortions produced by peculiar velocities and the Alcock-Paczynski effect using a theoretical modelling from \citet[]{Nadathur2019b}. The bias is treated as a nuisance parameter and the growth rate measurement is given in terms of $f\sigma_8$. In \citet[]{Hamaus2017,Achitouv2019, Hamaus2020}, the analysis performed on the void-galaxy cross-correlation provides a measurement in terms of $\beta$, using the RSD modelling from \citet[]{Cai2016}. In order to convert their measurements to a constraint on
$f\sigma_8$, we take the fiducial value
$b_1=1.85$~\citep{Alam2017} and compute $\sigma_8$ values for the Planck $\Lambda$CDM cosmology~\citep{Planck18}, giving $\sigma_8(z=0.32)=0.684$ and $\sigma_8(z=0.54)=0.612$. The corresponding $f\sigma_8$ constraints are $f\sigma_8(z=0.32)=0.757\pm0.17$ and $f\sigma_8(z=0.54)=0.517\pm0.063$ for \citet[]{Hamaus2017},  $f\sigma_8(z=0.32)=0.418\pm0.76$ and $f\sigma_8(z=0.54)=0.407\pm0.057$ for \citet[]{Achitouv2019}, and $f\sigma_8(z=0.51)=0.621\pm0.104$ for~\citet{Hamaus2020}.

The bottom panel of Figure~\ref{fig:fsig8} extends the comparison to other galaxy surveys: 6dFGS~\citep{Beutler2012}, WiggleZ~\citep{Blake2011}, VIPERS~\citep{Pezzotta2017} and FastSound~\citep{Okumura2016}. It is also interesting to compare our results to other measurements using voids, as in 6dFGS~\citep{Achitouv2017} and in VIPERS~\citep{Hawken2017}. We find a good consistency among all these measurements.

\begin{table}
    \caption{Final results on the growth rate estimate from the eBOSS DR16 void datasets. Mean values and errors on $\beta$ are taken from Table~\ref{table:data_ref}. The presented errors include the systematic component. The reported value of $b_1\sigma_8$ are taken from clustering analysis in the DR16 companion papers, for the LRG+CMASS sample~\citep{Bautista20,Gilmarin20}, the ELG sample~\citep{Tamone20,DeMattia20} and the QSO sample~\citep{Hou20,Neveux20}. The growth rate constraint results from applying Eq.~\ref{eqn:fsigma8} to these values. The total error quoted for $f\!\sigma_8$ includes the galaxy bias error contribution.}
    \centering
    \begin{tabular}{lcccc}
    \hline
    Data samples & $z_\textsf{eff}$ & $\beta $ & $b_1\sigma_8$ & $f\!\sigma_8$ \\
    \hline\hline
    LRG+CMASS &  0.740 & $0.415\pm0.087$ & $1.20\pm0.05$  &  $0.50\pm0.11$ \\
    ELG &  0.847 & $0.665\pm0.125$ & $0.78\pm0.05$  &  $0.52\pm0.10$ \\
    QSO &  1.478 & $0.313\pm0.134$ & $0.96\pm0.04$  &  $0.30\pm0.13$ \\
    \hline
    \hline
    \end{tabular}
    \label{table:result_fsig8}
\end{table}

\subsection{Discussion} 

The modelling of the redshift space distortions is undoubtedly our most important systematics (see Table~\ref{table:total_syst} and top panel of Figure~\ref{fig:fsig8}). This systematic effect is about the same order in the case of the ELGs and the LRGs, although the treatment of those mocks was different, \NSeries{} being cut sky mocks and \OR{} ELGs periodic boxes. For both LRGs and ELGs, systematic errors represent about $60\%$ of the statistical errors, and are therefore not the dominant errors in this analysis.

The most puzzling systematic appears in the analysis of the QSO sample for which we report a $40\%$ effect, more than twice the statistical uncertainty, dominated by the RSD modelling test. This effect is unexpected as RSD are supposed to be more compliant to linear theory at these epochs. 
However, it had already been reported that sampling density could have an impact on void properties, in particular for lower tracer density~\citep{Sutter2014b}.
For example, the sparsity of the QSO sample could lead to a dilution effect of the growth signal in the void-QSO cross-correlation function (see \citet{Cousinou2019}).
Two clues allow us to probe the role of sample density: firstly, systematics are well handled for denser samples such as LRGs and ELGs; secondly, we conducted a complementary study by subdividing the QSO \EZ{} sample according to the redshift, as described in Appendix \ref{sec:app_systQSO}. The comparison between recovered and fiducial $\beta$ values shows that the lower the QSO sample density, the higher the systematic error. This error becomes dominant for low void densities, at redshifts higher than z=1.9. 
There is a difference in the effect of the systematic bias between the \EZ{} QSO (27$\%$) and the \OR{} QSO (40$\%$). It might be attributed to additional degenerate effects such as the geometry (\EZ{}) or volume and statistics (\OR{}) of the samples.
The quoted systematic effect represents a conservative approach on the confidence of our measurement and accounts for the biases found in simulated data.

Nevertheless, our growth rate measurement in the DR16 QSO sample is found compatible at $1\sigma$ level with quasar clustering measurements and remains the first statistically significant measurement of the growth rate of structures at high redshift with voids found in this type of tracers. 

It is to be noted that the \NSeries{} redshift range is not the same as that of the LRG sample, but we consider that the mocks remain relevant to estimate the systematic effects of our modelling as they mimic the same tracer type. Future surveys and their preparatory phase (simulations and forecast) will enable us to investigate further these aspects.

Finally, such a systematic leads us to consider the RSD model itself. Our modelling, and subsequent analysis, relies on a ratio between the multipoles' contributions to measure the distortion parameter $\beta$, thus removing the dependence on the real-space correlation function. Such an implementation of the model might incur additional statistical fluctuations in the parameter determination.
This kind of systematics could probably be mitigated with added considerations such as the inclusion of velocity dispersion in the modelling or reconstruction of the real-space profile with additional nuisance parameters in the model as proposed in \citet{Hamaus2020}.


Recent papers \citep{Nadathur2019b,Nadathur2019} have also extended the modelling of the void-galaxy cross-correlation function further than the linear derivation of \citet{Cai2016}. This extended linear model is explicitly dependent on the real-space void-galaxy cross-correlation function and the real-space density profile of the void and their derivatives. But, as of today, the real-space density profile and the real-space correlation function of the void-galaxy are unknown theoretically and cannot,therefore, be predicted. This means that in order to obtain a constraint with the extended model, it is necessary to infer the real-space density profile from voids found in reconstructed galaxy samples or through empirical modelling. This requires an altogether different analysis than that presented in this paper as it correlates voids found in reconstructed galaxy samples with redshift-space galaxies. 
The extended model is very tuned to such an analysis and its main visible feature is a very different quadrupole behaviour from the model applied here. Said behaviour is not so noticeable in our analysis, which correlates redshift-space void and redshift-space galaxies, for several reasons: void centre definition, void finding, methodology choices or data noise. 
This model was applied to the DR16 LRG sample in \citet{Nadathur2020-eboss} and achieved tighter constraints by combining with galaxy clustering measurements as well as calibrating the true void density profile on simulations.

These models have also been applied to the same dataset in the past, in \citet{Hamaus2017,Achitouv2019} for the simple linear model, \citet{Hamaus2020} for the modified linear model and in \citet{Nadathur2019} for the extended model. These analyses obtained similar consistent constraints on the growth rate. For the sole purpose of constraining the growth rate of structure in the void-galaxy cross-correlation function in redshift space, we consider our modelling to be appropriate. 

Future galaxy surveys such as the Dark Energy Spectroscopic Instrument~\citep[DESI, ][]{desi2016a,desi2016b} and Euclid~\citep{Amendola2018} will tremendously increase the number of cosmic voids detected in the LSS and the statistical errors on their summary statistics. To fully benefit from this high statistical power, the systematic effects pertaining to void analysis, as shown in this work, need to be identified and thoroughly investigated.

\section{Summary and conclusions}
\label{sec:discuss}
In this paper we present the final void catalogues from the eBOSS DR16 datasets. We performed a multipole analysis in configuration space by computing the void-galaxy cross-correlation function for the three eBOSS tracers, the LRG, the ELG and the QSO samples, spanning a wide redshift range from $z=0.6$ to $z=2.2$. We have applied linear RSD modelling~\citep{Cai2016} to extract the distortion parameter and we have tested the validity of our approach using realistic N-body simulations.
We measured $\beta(z=0.74)=0.415\pm0.087$,  $\beta(z=0.85)=0.665\pm0.125$ and $\beta(z=1.48)=0.313\pm0.134$, for the CMASS+eBOSS LRG, the eBOSS ELG and the eBOSS QSO sample, respectively. 

In order to convert our measurements to a measurement of the growth rate $f\sigma_8$, we used consensus values of linear bias from the eBOSS DR16 companion papers~\citep{eBOSScosmo}, giving the following constraints: $f\sigma_8(z=0.74)=0.50\pm0.11$, $f\sigma_8(z=0.85)=0.52\pm0.10$ and $f\sigma_8(z=1.48)=0.30\pm0.13$. 

Voids have been predicted to be promising probes to constrain dark energy and modified gravity models. With the final data release DR16 of eBOSS, we have demonstrated that voids can be used as a competitive probe to constrain the growth rate of structure compared to that achieved with standard galaxy clustering. 
The clear improvement over our previous analysis using eBOSS DR14 data~\citep{Hawken2020} is due to the better statistics, since we have 2,800 voids and 4,300 voids in the DR16 catalogue as compared to 500 and 1,000 in the DR14 catalogue, for LRG and QSO sample respectively. In addition, we were able to create and use the ELG tracer catalogue, which contains almost 1,900 voids. 

Future spectroscopic galaxy surveys, such as DESI and Euclid, will observe between 35 and 50 million galaxies, and the consequent number of voids is expected to be more than 100,000, thrice that of the eBOSS sample.
The large amount of data will dramatically reduce statistical errors, both for conventional galaxy clustering analyses and for voids, and the challenge will be to keep systematic errors at the percent level. A new era of precision cosmology is emerging, which promises severe constraints on dark energy or modified gravity models. 

\section*{Data availability}
The data underlying this article were accessed from the SDSS Science Archive Server (\href{https://sas.sdss.org/}{https://sas.sdss.org/}). The derived data generated in this research will be shared on reasonable request to the corresponding author.

\section*{Acknowledgements}
This work was support by the eBOSS ANR grant (under contract ANR-16-CE31-0021) of the French National Research Agency. This work also acknowledges support from the OCEVU LABEX (Grant No. ANR-11-LABX-0060) and the A*MIDEX project (Grant No. ANR-11-IDEX-0001-02) funded by the Investissements d'Avenir french government program managed by the ANR.

SA is supported by the European Research Council through the COSFORM Research Grant (\#670193).

G.R. acknowledges support from the National Research Foundation of Korea (NRF) through Grants No. 2017R1E1A1A01077508 and No. 2020R1A2C1005655 funded by the Korean Ministry of Education, Science and Technology (MoEST).

Funding for the Sloan Digital Sky Survey IV has been provided by the Alfred P. Sloan Foundation, the U.S. Department of Energy Office of Science, and the Participating Institutions. SDSS-IV acknowledges support and resources from the centre for High-Performance Computing at the University of Utah. In addition, this research relied on resources provided to the eBOSS Collaboration by the National Energy Research Scientific Computing centre (NERSC). NERSC is a U.S. Department of Energy Office of Science
User Facility operated under Contract No. DE-AC02-05CH11231. The SDSS web site is www.sdss.org.

SDSS-IV is managed by the Astrophysical Research Consortium for the 
Participating Institutions of the SDSS Collaboration including the 
Brazilian Participation Group, the Carnegie Institution for Science, 
Carnegie Mellon University, the Chilean Participation Group, the French Participation Group, Harvard-Smithsonian centre for Astrophysics, 
Instituto de Astrof\'isica de Canarias, The Johns Hopkins University, Kavli Institute for the Physics and Mathematics of the Universe (IPMU) / 
University of Tokyo, the Korean Participation Group, Lawrence Berkeley National Laboratory, 
Leibniz Institut f\"ur Astrophysik Potsdam (AIP),  
Max-Planck-Institut f\"ur Astronomie (MPIA Heidelberg), 
Max-Planck-Institut f\"ur Astrophysik (MPA Garching), 
Max-Planck-Institut f\"ur Extraterrestrische Physik (MPE), 
National Astronomical Observatories of China, New Mexico State University, 
New York University, University of Notre Dame, 
Observat\'ario Nacional / MCTI, The Ohio State University, 
Pennsylvania State University, Shanghai Astronomical Observatory, 
United Kingdom Participation Group,
Universidad Nacional Aut\'onoma de M\'exico, University of Arizona, 
University of Colorado Boulder, University of Oxford, University of Portsmouth, 
University of Utah, University of Virginia, University of Washington, University of Wisconsin, Vanderbilt University, and Yale University.



\bibliographystyle{mnras}
\bibliography{biblio} 


\appendix
\section{Systematics in \EZ{} QSO}
\label{sec:app_systQSO}
To investigate the systematic effect identified in the \OR{} QSO, we focused on reproducing the analysis on the 1,000 QSO \EZ{}. 
First, we revisited the baseline analysis of the QSOs \EZ{} (see Section \ref{sec:ana_mock}) to estimate the shift of $ \langle \beta  \rangle$ from the expected $\beta^{\textsf{fid}}$. The latter is estimated from the expected growth rate and the bias measured in the standard galaxy clustering analyses \citep{Hou20, Neveux20}. The measured $\langle \beta \rangle$, the fiducial value, as well as the resulting deviation, are reported in Table \ref{table:app_mocks_ref}. The systematic measured in the QSO \EZ{} sample reduces to a $27\%$ effect.

In a second study, we considered the impact of the number density of objects on the recovery of the $\beta$ value. To this end, the QSO \EZ{} were separated into four equal volume redshifts bins. We then reproduced the analysis outlined in Section \ref{sec:ana_cor} and compared the recovered $\beta$ in each bin to the fiducial one $\beta_{\mathrm{fid}}$.   
$\beta_{\mathrm{fid}}$ was determined from the expected growth rate of the mocks and a fiducial bias inferred from the fitting function of the QSO bias given in \citep{duMasdesBourboux20} at the effective redshift of each bin as per eq. \ref{eq:eff_z}.
Table \ref{table:QSO_binning} lists, for each redshift range considered, the corresponding number density of both QSO and voids, both measured $\beta$ and expected $\beta_{\mathrm{fid}}$ value and the resulting deviation. The quantified systematic biases translate to the following relative effects, in order of increasing redshift bins :
\begin{eqnarray}
\label{eqn:syst_zbins}
\sigma_{\textsf{syst,zbins}} = (-27.3\%, -13.5\%, -19.7\%, -63.4\%).
\end{eqnarray}
The less biased bins are the most central ones, where the number density of both voids and galaxies is greater. The outer bins are more affected by systematic effects, especially the higher redshift one. The latter is significantly sparse in comparison, leading to a $63\%$ effect. We can therefore conclude that there is a correlation between the systematic bias on the measured $\langle \beta \rangle$ and the number density of the objects.

\begin{table}
    \centering
    \caption{Recovered $\langle \beta \rangle$ from the 1,000 \EZ{} QSO quoted with its associated error $\langle \sigma_\beta \rangle$, the mean of the individual fitting errors. $\beta_{\mathrm{fid}}$ is estimated from the expected fiducial growth rate and the bias recovered from the standard galaxy clustering analysis of \citet{Hou20,Neveux20}. The last column quantifies the bias from the expected $\beta_{\mathrm{fid}}$ along with its statistical error ($\frac{\langle \sigma_\beta \rangle}{\sqrt{N}}$).}
    \begin{tabular}{cccc}
    \hline
    EZmocks & $\langle \beta \rangle \pm \langle \sigma_\beta \rangle$ &  $\beta^{\textsf{fid}}$&  $ \langle \beta \rangle - \beta_{\mathrm{fid}} $ \\
    \hline\hline
    QSO & 0.294 $\pm$ 0.049  & 0.403 &  $-0.109 \pm 0.002$ \\
    \hline
    \hline
    \end{tabular}
    \label{table:app_mocks_ref}
\end{table}

\begin{table}
    \centering
    \caption{Resulting systematic bias for each equal volume redshift bin from the 1,000 \EZ{} QSO. $\langle \beta \rangle$ and $\langle \sigma_\beta \rangle$ are the mean of the individual fitting values and errors, respectively. $z_{\mathrm{eff}}$ is the effective redshift at in each bin at which the fiducial $\beta_{\mathrm{fid}}$ was evaluated. The latter is estimated from the expected fiducial growth rate and the empirical QSO bias given in \citet{duMasdesBourboux20}. $\bar{n}_v$ and $\bar{n}_g$ are the average number density of voids and QSOs (respectively). The last column quantifies the bias from the expected $\beta_{\mathrm{fid}}$ along with its statistical error ($\frac{\langle \sigma_\beta \rangle}{\sqrt{N}}$). }
    \begin{tabular}{lcccccc}
        \hline \hline
        z-Bin & $z_{\mathrm{eff}}$ & $\bar{n}_g | \bar{n}_v$ & $\langle \beta \rangle \pm \langle \sigma_\beta \rangle$ & $\beta_{\mathrm{fid}}$ & $\langle \beta \rangle - \beta_{\mathrm{fid}}$  \\
        & & $\times 10^{-5} | \times 10^{-7}$ & & &\\
        
        \hline \hline
        $[0.8, 1.24]$ & 1.06 & $1.478 |  2.077$  &$0.356 \pm 0.111$ & 0.49 &  $-0.134 \pm 0.004$ \\ 
        $[1.24, 1.58]$ & 1.41 &  $1.556  |  2.477$  &$0.353 \pm 0.093$ & 0.41 & $-0.057 \pm 0.003$ \\
        $[1.58, 1.9]$ &  1.74   & $1.386  |  2.244$  &$0.280 \pm 0.098$ & 0.35 &  $-0.070 \pm 0.003$ \\ 
        $[1.9, 2.2]$ & 2.00  & $1.047  | 1.386$  &$0.113 \pm 0.152$ & 0.31 &  $-0.197 \pm 0.005$ \\ 
        \hline \hline
    \end{tabular}
    \label{table:QSO_binning}
\end{table}

\bsp	
\label{lastpage}
\end{document}